\documentclass[pdflatex,sn-nature,referee,a4paper]{sn-jnl}

\usepackage{graphicx}%
\usepackage{multirow}%
\usepackage{amsmath,amssymb,amsfonts}%
\usepackage{placeins}

\usepackage{amsthm}%
\usepackage{mathrsfs}%
\usepackage[title]{appendix}%
\usepackage{xcolor}%
\usepackage{textcomp}%
\usepackage{manyfoot}%
\usepackage{booktabs}%
\usepackage{algorithm}%
\usepackage{algorithmicx}%
\usepackage{algpseudocode}%
\usepackage{listings}%
\usepackage{anyfontsize}%
\usepackage[separate-uncertainty=true,multi-part-units=single]{siunitx}%
\usepackage{physics}%
\usepackage{xspace}%

\DeclareSIUnit{\sample}{Sa}

\newcommand{\muLL}{\mu_\mathrm{L1}}

\newcommand{\muLR}{\mu_\mathrm{L2}}
\newcommand{\muC}{\mu_\mathrm{C}}
\newcommand{\muRL}{\mu_\mathrm{R1}}

\newcommand{\muRR}{\mu_\mathrm{R2}}

\newcommand{\tL}{t_\mathrm{L}}
\newcommand{\DeltaL}{\Delta_\mathrm{L}}
\newcommand{\tR}{t_\mathrm{R}}
\newcommand{\DeltaR}{\Delta_\mathrm{R}}

\newcommand{\tLC}{t_\mathrm{LC}}
\newcommand{\tCR}{t_\mathrm{CR}}
\newcommand{\tLR}{t_\mathrm{LR}}
\newcommand{\epsL}{\epsilon_\mathrm{L}}
\newcommand{\epsR}{\epsilon_\mathrm{R}}
\newcommand{\lambdaL}{\lambda_\mathrm{L}}
\newcommand{\lambdaR}{\lambda_\mathrm{R}}

\newcommand{\VLL}{V_\mathrm{L1}}
\newcommand{\VLH}{V_\mathrm{LH}}
\newcommand{\VLR}{V_\mathrm{L2}}
\newcommand{\VC}{V_\mathrm{C}}
\newcommand{\VRL}{V_\mathrm{R1}}
\newcommand{\VRH}{V_\mathrm{RH}}
\newcommand{\VRR}{V_\mathrm{R2}}
\newcommand{\VPS}{V_\mathrm{PS}}
\newcommand{\QDLL}{QDL1\xspace}
\newcommand{\QDLR}{QDL2\xspace}
\newcommand{\QDC}{QDC\xspace}
\newcommand{\QDRL}{QDR1\xspace}
\newcommand{\QDRR}{QDR2\xspace}
\newcommand{\QDPS}{QDPS\xspace}

\newcommand{\SM}{S_\mathrm{M}^\mathrm{11}}

\newcommand{\peeee}{p\mathrm{(ee|ee)}}
\newcommand{\poooo}{p\mathrm{(oo|oo)}}
\newcommand{\peeoo}{p\mathrm{(ee|oo)}}
\newcommand{\pooee}{p\mathrm{(oo|ee)}}

\newcommand{\peoeo}{p\mathrm{(eo|eo)}}
\newcommand{\poeoe}{p\mathrm{(oe|oe)}}
\newcommand{\peooe}{p\mathrm{(eo|oe)}}
\newcommand{\poeeo}{p\mathrm{(oe|eo)}}

\newcommand{\pee}{p\mathrm{(e|e)}}
\newcommand{\peo}{p\mathrm{(e|o)}}
\newcommand{\poe}{p\mathrm{(o|e)}}
\newcommand{\poo}{p\mathrm{(o|o)}}

\begin{document}

\title[Article Title]{
Coherent parity oscillations in coupled minimal Kitaev chains\newline

}

\author[1]{\fnm{Francesco} \sur{Zatelli}}
\equalcont{These authors contributed equally to this work.}

\author[1]{\fnm{Bart} \sur{Roovers}}
\equalcont{These authors contributed equally to this work.}

\author*[1]{\fnm{Nick} \spfx{van} \sur{Loo}}
\email{n.vanloo@tudelft.nl}
\equalcont{These authors contributed equally to this work.}

\author[1]{\fnm{Antonio} \sur{Lombardi}}
\author[1]{\fnm{Juan D.} \sur{Torres Luna}}
\author[1]{\fnm{Sebastian} \sur{Miles}}
\author[1]{\fnm{Vincent P.M.} \sur{Sietses}}
\author[1]{\fnm{Florian J.} \sur{Bennebroek Evertsz'}}
\author[1]{\fnm{Pablo} \sur{Cova Fariña}}
\author[1]{\fnm{Alberto} \sur{Bordin}}

\author[2]{\fnm{Ghada} \sur{Badawy}}
\author[2]{\fnm{Erik P.A.M.} \sur{Bakkers}}
\author[1]{\fnm{Michael} \sur{Wimmer}}
\author*[1]{\fnm{Leo P.} \sur{Kouwenhoven}}
\email{l.p.kouwenhoven@tudelft.nl}

\affil[1]{\orgdiv{QuTech and Kavli Institute of Nanoscience}, \orgname{Delft University of Technology},
\country{The Netherlands}}

\affil[2]{\orgdiv{Department of Applied Physics}, \orgname{Eindhoven University of Technology},

\country{The Netherlands}}


\abstract{
    Majorana zero modes provide a route to fault-tolerant qubits by encoding information non-locally in fermion parity. Their sensitivity to noise is expected to decrease exponentially with increasing separation between the Majoranas, a suppression known as topological protection~\cite{kitaev_unpaired_2001}. 
    Kitaev chains engineered in quantum dot--superconductor arrays provide a tunable platform in which separated Majorana zero modes can emerge at the ends of the chain~\cite{kitaev_unpaired_2001, sau_realizing_2012}, even in two-site chains~\cite{leijnse_parity_2012}. These minimal-chain modes are known as poor man’s Majoranas and retain characteristic Majorana properties, including near-zero energy and equal electron--hole character, but have only limited protection~\cite{leijnse_parity_2012,dvir_realization_2023,ten_haaf_two-site_2024}. 
    A key outstanding challenge is to move beyond identifying such modes in electrical transport measurements and achieve coherent qubit control in the time domain.
    Here, we demonstrate a Majorana parity qubit by realizing coherent coupling between two-site Kitaev chains~\cite{leijnse_parity_2012, tsintzis_majorana_2024, pino_minimal_2024, pan_rabi_2025}. 
    Since total fermion parity is conserved, the system separates into global even and odd parity manifolds. We observe coherent parity oscillations in both manifolds with equal oscillation frequencies at the Majorana sweet spot, as predicted for isolated Majorana zero modes~\cite{tsintzis_majorana_2024}. We further show that the oscillation frequency and coherence depend systematically on inter-chain coupling and quantum-dot detunings, in close agreement with our model for short, partially protected chains.
    Our results establish the first coherent control of a Majorana qubit, encoded in the fermion parity of Majorana zero modes in minimal Kitaev chains.
}

\maketitle
\pagebreak
Two Majorana modes form a non-local fermion, corresponding to two degenerate states with even ($\ket{e}$) and odd ($\ket{o}$) parity~\cite{kitaev_unpaired_2001}. Two pairs of Majorana modes give rise to four parity states $\ket{ee}$, $\ket{oo}$, $\ket{eo}$, and $\ket{oe}$. The first two states ($\ket{ee}$ and $\ket{oo}$) span the global even parity manifold, whereas the latter two ($\ket{eo}$ and $\ket{oe}$) span the global odd parity manifold. Transitions between these global manifolds require adding or removing a fermion via exchange with an external source, and are therefore typically incoherent processes~\cite{rainis_majorana_2012}. For this reason, qubits defined with Majorana modes are usually restricted to either the global even or global odd parity sector~\cite{bravyi_fermionic_2002,nayak_non-abelian_2008}.

\begin{figure*}[h!]
    \centering
    \includegraphics[width=1\textwidth]{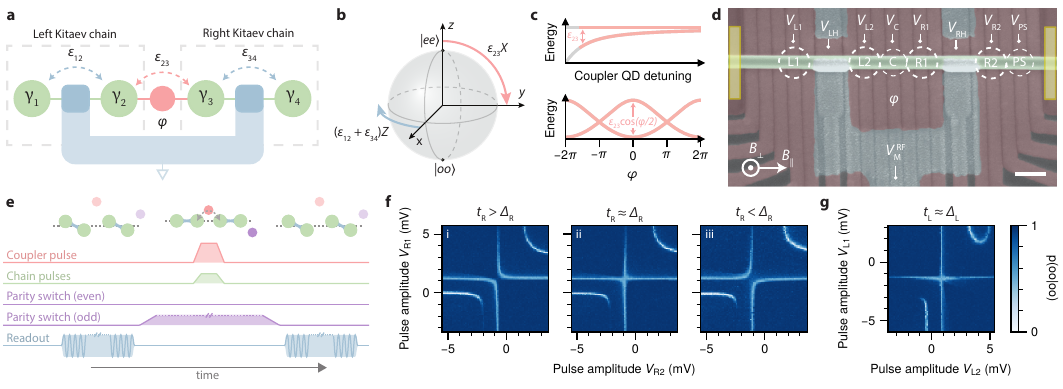}
    \caption{\textbf{Tuning a Majorana parity qubit.}
    \textbf{a,} Schematic of a Majorana qubit formed by two minimal Kitaev chains. Four Majorana modes ($\gamma_1,\gamma_2,\gamma_3,\gamma_4$) encode the qubit, while a central quantum dot (red) mediates inter-chain coupling. A superconducting segment (blue) fixes the superconducting phase $\varphi$. 
    \textbf{b,} Bloch-sphere representation of the qubit in the global even parity manifold. The poles correspond to $\ket{ee}$ and $\ket{oo}$. Rotations are generated by tuning the pairwise Majorana couplings $\varepsilon_{ab}$.
    \textbf{c,} Example energy spectra of the computational subspace versus coupler-dot detuning at $\varphi=0$ (top) and as a function of $\varphi$ at finite coupling (bottom). The states are nearly degenerate at large detuning, while decreasing detuning increases their splitting $\varepsilon_{23}$. The splitting oscillates with $\varphi$ and vanishes at $\varphi=\pi$.
    \textbf{d,} False-colored scanning electron micrograph of the device with the InSb nanowire (green), Al superconductor (blue) and Ti/Pd bottom gates (red). Gate voltages define quantum dots in the nanowire (white dashed circles). Yellow boxes indicate Cr/Au normal contacts fabricated after imaging. Magnetic field orientations and superconducting phase difference $\varphi$ are indicated. Scale bar, \SI{200}{\nano m}.
    \textbf{e,} A readout–control–readout cycle is implemented by a pulse sequence. For readout, the chains are detuned and the coupler QD is placed far above resonance. Parity states are measured dispersively using a radio-frequency resonator coupled to the superconductor, whose resonance frequency shifts due to the parity-dependent quantum capacitance of the device. During control, the chain QDs are pulsed near charge degeneracy and the coupler QD is brought closer to resonance. All pulse amplitudes are applied in virtualized gate coordinates that compensate for cross-capacitance between the relevant gates.
    To access the odd parity manifold, we pulse the parity-switching dot through resonance, flipping the parity of the right chain: $\ket{ee}\leftrightarrow\ket{eo}$ and $\ket{oo}\leftrightarrow\ket{oe}$.
    \textbf{f,} Pulsed charge-stability diagrams showing $\poooo$ as a function of the right-chain plunger gates ($\VRL$,$\VRR$) for different values of the hybrid gate $\VRH$. Varying $\VRH$ changes the relative strength of tunneling and pairing in the chain, driving a crossover from $\tR>\DeltaR$ (i), through the Majorana sweet spot $\tR\approx\DeltaR$ (ii), to $\tR<\DeltaR$ (iii). In addition to the ground state parity transitions, excited states of the chain are visible.
    \textbf{g,} $\poooo$ as a function of the left-chain plunger gates ($\VLL$,$\VLR$), showing the Majorana sweet spot ($\tL\approx\DeltaL$) of the left chain.
    }
    \label{fig:1}
\end{figure*}

Transitions within a global parity manifold, i.e. between $\ket{ee}$ and $\ket{oo}$, or between $\ket{eo}$ and $\ket{oe}$, instead involve an internal exchange of a fermion between different Majorana pairs, which can be a coherent process. As a result, Majorana-based qubits are commonly encoded using four Majorana modes, denoted $\gamma_1$, $\gamma_2$, $\gamma_3$, and $\gamma_4$. The parity of a pair of Majorana modes $\gamma_a,\gamma_b$ is defined as $i\gamma_a\gamma_b$.  A coupling between two Majorana modes is described by $i\varepsilon_{ab}\gamma_a\gamma_b$, where $\varepsilon_{ab}$ is the corresponding coupling strength~\cite{sarma_majorana_2015,tsintzis_majorana_2024,pino_minimal_2024, pan_rabi_2025}.
Because the four Majoranas are realized in two Kitaev chains, with $\gamma_1,\gamma_2$ on the left and $\gamma_3,\gamma_4$ on the right (Fig.~\ref{fig:1}a), the individual chain parities provide the natural computational basis. We therefore define the Pauli operators in the global even manifold as $Z=i\gamma_1\gamma_2=i\gamma_3\gamma_4$ and $X=i\gamma_2\gamma_3=i\gamma_1\gamma_4$, where the latter describes coupling between the chains. Assuming only nearest-neighbor couplings, the effective qubit Hamiltonian in the even manifold is

\begin{equation}  \label{eq:hamiltonian_even}
    H^{\mathrm{even}} = (\varepsilon_{12}+\varepsilon_{34}) Z + \varepsilon_{23} \cos(\varphi/2)X,
\end{equation}
where $\varphi$ is the superconducting phase difference between the two chains~\cite{kitaev_unpaired_2001,fu_josephson_2009,tsintzis_majorana_2024, pino_minimal_2024}.
The first term is longitudinal in the logical $Z$ basis and sets the energy difference between the computational states, while the second term is transverse and mixes the two states~\cite{ivanov_non-abelian_2001,bravyi_fermionic_2002,sarma_majorana_2015,tsintzis_majorana_2024,pino_minimal_2024, pan_rabi_2025}.
By tuning these couplings, arbitrary single-qubit rotations can be generated~\cite{sarma_majorana_2015,tsintzis_majorana_2024,pan_rabi_2025}. The same qubit can also be operated in the odd manifold with computational basis $\ket{eo}$, $\ket{oe}$ and effective Hamiltonian
\begin{equation} \label{eq:hamiltonian_odd}
H^{\mathrm{odd}} = (\varepsilon_{12}-\varepsilon_{34}) Z + \varepsilon_{23}\cos(\varphi/2) X.
\end{equation}
The sign change reflects the opposite chain parities in the odd manifold. 
When $\varepsilon_{12}=\varepsilon_{34}=0$, the inter-chain coupling $\varepsilon_{23}$ produces identical oscillation frequencies in the two manifolds, while finite $\varepsilon_{12}, \varepsilon_{34}$ can make the two frequencies differ.
Imperfect Majorana localization can also make the qubit frequency parity-dependent: if the outer Majorana modes $\gamma_1$ and $\gamma_4$ delocalize to the inner QDs, they can generate an effective $i \varepsilon_{14}\gamma_1 \gamma_4$ coupling, which contributes to $X$ with opposite sign in the two manifolds. Similarly, hybridization of $\gamma_1$ with $\gamma_3$, or $\gamma_2$ with $\gamma_4$, generates components along $Y$ (see Methods section~\ref{methods:effective_full_hamiltonian})~\cite{tsintzis_majorana_2024}. Any discrepancy between the oscillation frequencies  in the global even and odd manifolds therefore provides a measure of deviations from ideal Majoranas~\cite{tsintzis_majorana_2024,luethi_perfect_2024}.

\section*{Tuning a qubit device}\label{Tuning}

We realize the Majorana parity qubit described above in an InSb nanowire~\cite{badawy_high_2019} containing two minimal Kitaev chains (Fig.~\ref{fig:1}d). Bottom gates tune the left-chain QDs (\QDLL, \QDLR with $\VLL,\VLR$), the right-chain QDs (\QDRL, \QDRR with $\VRL,\VRR$), a central QD acting as a tunable coupler (\QDC with $\VC$)~\cite{leijnse_parity_2012,tsintzis_majorana_2024}, and an auxiliary parity-switching dot (\QDPS with $\VPS$)~\cite{flensberg_non-abelian_2011}. Between the QDs within each chain, nanowire segments proximitized by Al strips host Andreev bound states~\cite{deacon_tunneling_2010, pillet_andreev_2010,lee_spin-resolved_2014} that mediate elastic co-tunneling and crossed-Andreev reflection, with amplitudes $\tL,\DeltaL$ and $\tR,\DeltaR$ tuned by $\VLH$ and $\VRH$~\cite{recher_andreev_2001,hofstetter_cooper_2009,liu_tunable_2022,dvir_realization_2023,bordin_tunable_2023}. The Al strips form a loop for phase control. Single-shot parity readout is performed through quantum capacitance using an off-chip resonator~\cite{liu_fusion_2023,steiner_readout_2020,van_loo_single-shot_2026,pan_rabi_2025,zhang_gate_2026}, and fast baseband pulses are applied through bias tees. Measurements are performed at $B_{||}=\SI{150}{\milli T}$ and base temperature $\approx\SI{20}{\milli K}$. Further setup details and cross-capacitance compensation are described in Fig.~\ref{fig:ed_measurementsetup} and Methods sections~\ref{Methods_setup} and~\ref{Methods_pulses}.

In minimal Kitaev chains, poor man's Majoranas appear at a fine-tuned sweet spot where $\tL = \DeltaL$ and $\tR = \DeltaR$~\cite{leijnse_parity_2012,liu_tunable_2022,dvir_realization_2023, tsintzis_creating_2022, luethi_perfect_2024}. We first approach this regime using transport measurements for coarse tuning~\cite{zatelli_robust_2024}. We then pinch off the cutter gates next to the normal leads to fix total parity.
In this isolated regime the parity lifetime is long enough for single-shot parity measurement.
Our readout then distinguishes $\ket{ee}$ from $\ket{oo}$ and detects leakage into the global odd sector, but does not directly resolve $\ket{eo}$ from $\ket{oe}$ (Fig.~\ref{fig:ed_readout})~\cite{van_loo_single-shot_2026,pan_rabi_2025}. 
For operations in the odd manifold, we use the parity-switching dot to map odd states to the even manifold for readout (see Methods section~\ref{Methods_pulses}).

To fine-tune the chains in the isolated regime, we couple \QDC to the adjacent QDs (Fig.~\ref{fig:ed_rampduration}) and measure charge-stability diagrams with the pulse sequence described in Fig.~\ref{fig:1}e and Methods section~\ref{Methods_pulses}. Each sequence starts with parity readout in a detuned configuration, where the coupler QD is far off resonance and the chain QDs are detuned from charge degeneracy to suppress inter-chain hopping (Fig.~\ref{fig:ed_readout}). The chain QDs are then pulsed to charge degeneracy with a ramp time of \SI{25}{ns}. Simultaneously, the coupler QD is brought closer to resonance, mediating an effective hybridization between the chains and enabling parity mixing. After \SI{60}{ns}, the system is pulsed back to the detuned configuration for a final readout. Repeating the sequence yields the probabilities of the final parity states conditioned on the initial state.

Fig.~\ref{fig:1}f shows the result of this protocol for initialization in $\ket{oo}$, while varying the pulse amplitudes applied to the right-chain QDs, with the left chain near its sweet spot ($\varepsilon_{12} \approx 0$).
We interpret the observed lines as degeneracies between global even states of the chains, where inter-chain coupling enables parity mixing.
The inner lines mark degeneracies between $\ket{ee}$ and $\ket{oo}$, corresponding to $\varepsilon_{34}\approx0$ (see Eq.~\eqref{eq:hamiltonian_even}).  
The outer lines in the top-right and bottom-left quadrants arise when the energy of the right chain is aligned with the first excited state of the left chain (see Fig.~\ref{fig:ed_PCSDs_excited} and Methods section~\ref{Methods_pulsedCSD}).
Repeating the measurement for different values of $\VRH$ shows that the system can be tuned from a regime with an avoided crossing indicating $\tR>\DeltaR$ (Fig.~\ref{fig:1}f, i), through a crossing where $\tR \approx \DeltaR$ (Fig.~\ref{fig:1}f, ii), and into the opposite regime with $\tR < \DeltaR$ (Fig.~\ref{fig:1}f, iii)~\cite{dvir_realization_2023}. An analogous procedure is used to tune the left chain, with the right chain held near its sweet spot (Fig.~\ref{fig:1}g). As discussed in Methods section~\ref{Methods_tuneup} and Fig.~\ref{fig:ed_PCSDs}, the final tuning is verified by confirming that the same crossings are also observed in the global odd parity manifold.

Details of the tune-up and pulsed charge-stability measurements are provided in Methods section~\ref{Methods_tuneup} and~\ref{Methods_pulsedCSD}, respectively. Supporting measurements, including the predicted shift of the degeneracy lines when either chain is detuned from its sweet spot ($\varepsilon_{12}\neq 0$, $\varepsilon_{34}\neq 0$), analogous data for the global odd manifold, and additional maps for different $\VC$ pulse amplitudes, are shown in Fig.~\ref{fig:ed_PCSDs}, Fig.~\ref{fig:ed_PCSDs_excited}, and Fig.~\ref{fig:ed_pulsedCSDsQDC}.

\section*{Coherent parity oscillations}\label{oscillations}
\begin{figure*}[h!]
    \centering
    \includegraphics[width=\textwidth]{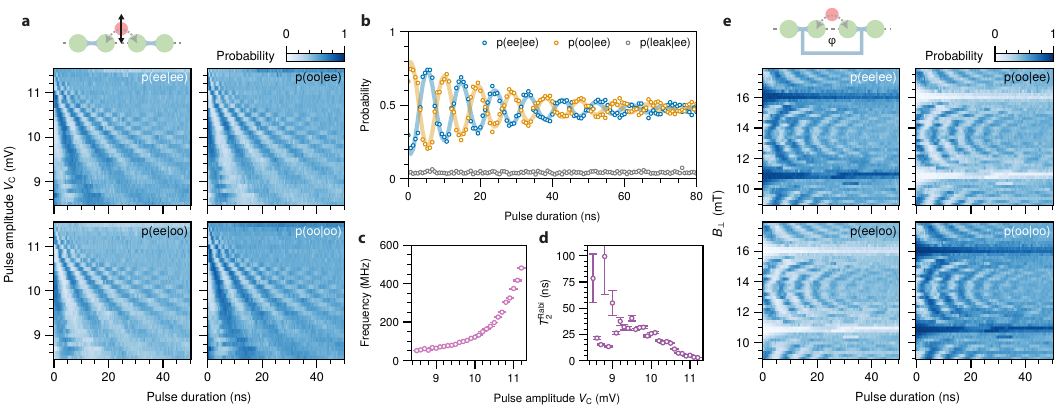}
    \caption{\textbf{Coherent control and phase dependence of the Majorana qubit.}
    \textbf{a,} Conditional probabilities $\peeee$, $\pooee$, $\peeoo$, and $\poooo$ as a function of coupling pulse duration and amplitude $\VC$, showing coherent oscillations between the qubit states.  
    \textbf{b,} $\peeee$, $\pooee$, and leakage to the odd-parity manifold as a function of pulse duration at $\VC = \SI{9.8}{mV}$.
    \textbf{c,d,} Oscillation frequency (\textbf{c}) and Rabi decay time $T_2^{\mathrm{Rabi}}$ (\textbf{d}) extracted from panel \textbf{a} as a function of $\VC$.
    \textbf{e,} $\peeee$, $\pooee$, $\peeoo$, and $\poooo$ as a function of pulse duration and out-of-plane magnetic field $B_\perp$ which controls $\varphi$. The oscillation frequency is strongly suppressed near $\varphi=\pi$, where inter-chain coupling vanishes.
    }
    \label{fig:2}
\end{figure*}
The pulsed charge-stability diagrams (Fig.~\ref{fig:1}f,g) show parity mixing within \SI{60}{\nano s}. To resolve the details of these fast dynamics, we pulse both chains to the Majorana sweet spot and vary the pulse duration and amplitude applied to the coupler QD. 
Fig.~\ref{fig:2}a shows the resulting conditional probabilities for the two initial states in the even parity manifold. The chain parities oscillate between $\ket{ee}$ and $\ket{oo}$, with a frequency that increases for larger $\VC$ pulse amplitude as the coupler QD is brought closer to resonance. We interpret these oscillations as Rabi oscillations between parity states of the coupled chains. Near $\SI{11.2}{\milli V}$, the oscillations become too fast to resolve.
The large oscillation contrast, even at relatively low frequency, is consistent with $\ket{ee}$ and $\ket{oo}$ being nearly degenerate compared with the inter-chain coupling that drives the oscillations ($\varepsilon_{12}+\varepsilon_{34}\ll \varepsilon_{23}$).
Fig.~\ref{fig:2}b shows a separate measurement of the oscillations for initialization in $\ket{ee}$, taken at a $\VC$ pulse amplitude of \SI{9.8}{\milli V}. The population remains within the even-parity manifold, with leakage stable around 5\%, which we attribute to readout errors from limited signal-to-noise ratio or residual parity switching during measurement. We fit the data to a damped oscillation with an exponential envelope and extract a decay time of $T_2^{\mathrm{Rabi}} = \SI{26(2)}{\nano s}$ for an oscillation frequency of $f = \SI{106.7(0.4)}{\mega Hz}$.

Fitting each line in Fig.~\ref{fig:2}a where oscillations are resolved shows that the frequency is inversely proportional to the detuning of \QDC from resonance (Fig.~\ref{fig:2}c and Fig.~\ref{fig:ed_detunings}m), consistent with coupler-mediated elastic co-tunneling~\cite{leijnse_parity_2012,tsintzis_majorana_2024}. The Rabi decay time is non-monotonic in $\VC$ (Fig.~\ref{fig:2}d). We interpret this as a crossover between two dephasing regimes.
For strong coupling, i.e. fast oscillations, charge noise on the coupler QD dominates dephasing, as the charge dispersion increases near resonance (Fig.~\ref{fig:1}c). 
Farther from resonance, the oscillations slow down as the effective inter-chain coupling is reduced. In this regime, dephasing is dominated by noise on the chains, due to the reduced protection provided by the coupling~\cite{pan_rabi_2025}.
Additional dephasing may also arise from the readout circuit.

After demonstrating electrical control of the inter-chain coupling through \QDC, we probe its predicted dependence on the superconducting phase difference. We fix the $\VC$ pulse amplitude and vary the out-of-plane magnetic field $B_\perp$ (Fig.~\ref{fig:2}e). We observe a periodic dependence of the oscillation frequency, with a period of $\approx \SI{5}{\milli T}$, consistent with the loop area (approximately $\SI{0.4}{\micro m^2}$). The oscillation frequency is suppressed on approaching $B_\perp \approx \SI{11}{\milli T}$ and $B_\perp \approx \SI{16}{\milli T}$ from either side. At these fields, we observe neither oscillations nor fast dephasing. This observation is consistent with the theoretical prediction that the effective coupling between the two Kitaev chains vanishes at $\varphi = \pi$ (see Eq.~\eqref{eq:hamiltonian_even})~\cite{alicea_non-abelian_2011, tsintzis_majorana_2024}. In the manuscript, measurements are performed at $\varphi \approx 0$ unless specified otherwise. 

\section*{Comparing even and odd parity oscillations}\label{oddmanifold}

\begin{figure*}[h!]
    \centering
    \includegraphics[width=1\textwidth]{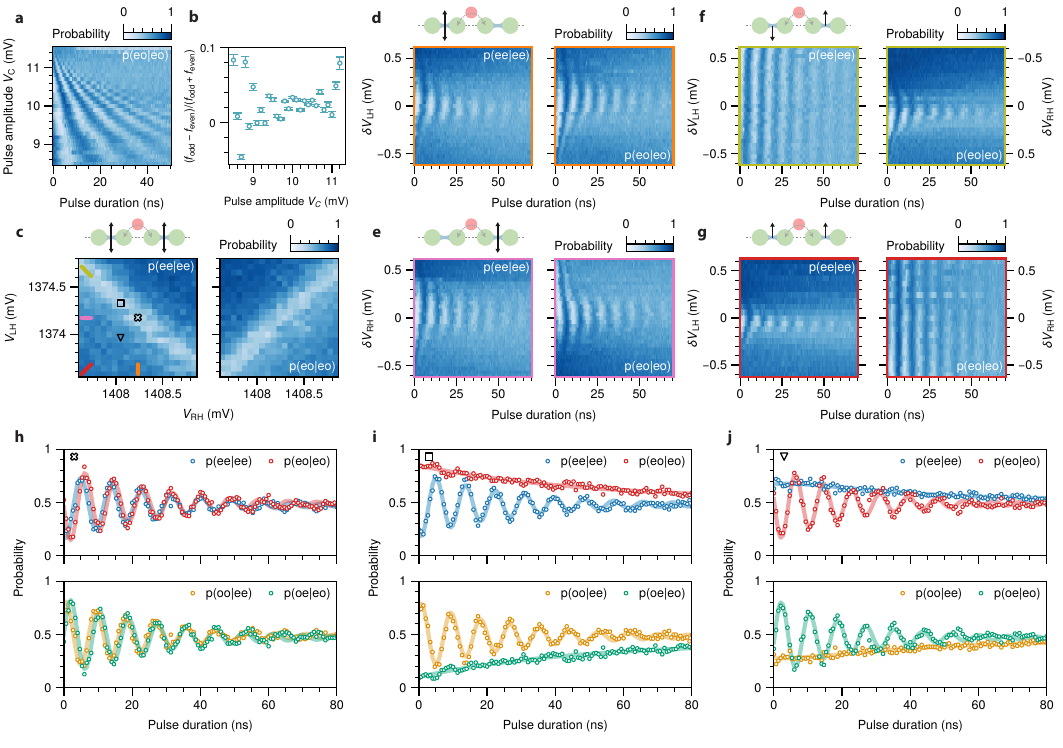}
    \caption{\textbf{Comparison between even and odd global parity manifolds.}
    \textbf{a,} $\peoeo$ as a function of pulse duration and amplitude $\VC$, showing coherent oscillations in the odd manifold.  
    \textbf{b,} Relative frequency difference between the even- and odd-manifold oscillations as a function of $\VC$. 
    \textbf{c,}  $\peeee$ (left) and $\peoeo$ (right) as a function of the left and right hybrid-gate voltages ($\VLH$ and $\VRH$) for a fixed pulse duration of \SI{18}{ns}. Parity mixing occurs along diagonal or anti-diagonal lines depending on the global parity manifold. Their intersection (cross) identifies the Majorana sweet spot, where the resonance conditions of the even and odd manifolds coincide. The colored marks indicate sweep directions for panels \textbf{d-g}.
    \textbf{d,e,} Horizontal and vertical cuts through the Majorana sweet spot, obtained by varying $\VLH$ (\textbf{d}) and $\VRH$ (\textbf{e}). Oscillations are observed in both parity manifolds near the center and vanish simultaneously upon detuning.
    \textbf{f,g,} Anti-diagonal (\textbf{f}) and diagonal (\textbf{g}) cuts through the Majorana sweet spot. Along the anti-diagonal direction, oscillations are observed over the full scan range in the even manifold and only near the center in the odd manifold. Along the diagonal direction, the roles of the two manifolds are reversed.  
    \textbf{h,} Pulse-duration dependence at the Majorana sweet spot, where coherent oscillations are observed with nearly equal frequencies ($f_\mathrm{even} = \SI{113.4\pm0.5}{MHz}$, $f_\mathrm{odd} = \SI{117.7\pm0.4}{MHz}$) in both parity manifolds.
    \textbf{i,j,} Pulse-duration dependence at the square and triangle locations in \textbf{c}, corresponding to anti-diagonal (\textbf{i}) and diagonal (\textbf{j}) detuning. In these configurations, coherent oscillations are observed in only one of the parity manifolds.
    }
    \label{fig:3}
\end{figure*}

While the previous measurements establish degeneracy between $\ket{ee}$ and $\ket{oo}$, they do not show that the two individual chains are simultaneously degenerate. For example, equal splittings of the two chains with opposite signs would cancel in the even manifold while adding in the odd manifold (see Eqs.~\eqref{eq:hamiltonian_even} and~\eqref{eq:hamiltonian_odd}).
In general, for isolated Majorana zero modes, the inter-chain coupling is predicted to generate the same Rabi frequency in the even and odd parity sectors. 
Comparing the coherent dynamics in both manifolds therefore tests whether the individual chains are both at the Majorana sweet spot and can reveal deviations from the ideal limit due to wavefunction overlap (see Methods section~\ref{methods:consequences_imperfect_majoranas})~\cite{tsintzis_majorana_2024}.
 
Fig.~\ref{fig:3}a shows parity oscillations in the odd manifold as a function of pulse duration and amplitude. The extracted oscillation frequencies typically agree with those measured in the even manifold to about 5\% (Fig.~\ref{fig:3}b).
To test the behavior away from the Majorana sweet spot, we detune the chains by varying $\VLH$ and $\VRH$, i.e. making $\tL\neq \DeltaL$ and $\tR \neq \DeltaR$, at fixed pulse duration ($\SI{18}{\nano s}$) and $\VC$ pulse amplitude (\SI{9.8}{\milli V}) (Fig.~\ref{fig:3}c). Enhanced parity mixing appears along the antidiagonal in the even manifold and along the diagonal in the odd manifold. Time-domain measurements show that detuning either hybrid gate individually suppresses the oscillation contrast and coherence time in both manifolds, while slightly increasing the oscillation frequency  (Fig.~\ref{fig:3}d,e). By contrast, simultaneous antisymmetric detuning preserves oscillations only in the even manifold (Fig.~\ref{fig:3}f) and symmetric detuning only in the odd manifold (Fig.~\ref{fig:3}g).
Data for the other initial states are available in Fig.~\ref{fig:ed_detunings} and Fig.~\ref{fig:ed_detunings_odd}. 
Separate representative measurements further illustrate these three regimes: oscillations in both manifolds at the Majorana sweet spot (Fig.~\ref{fig:3}h), oscillations only in the even manifold for antisymmetric detuning (Fig.~\ref{fig:3}i), and oscillations only in the odd manifold for symmetric detuning (Fig.~\ref{fig:3}j).

These observations are consistent with the parity-dependent form of the qubit Hamiltonian (Eqs.~\eqref{eq:hamiltonian_even} and~\eqref{eq:hamiltonian_odd}). Detuning the hybrid gates generates intra-chain splittings $\varepsilon_{12}=(\tL-\DeltaL)/2$ and $\varepsilon_{34}=(\tR-\DeltaR)/2$, making the longitudinal term in general non-zero: $\varepsilon_{12}+\varepsilon_{34}$ in the even manifold and $\varepsilon_{12}-\varepsilon_{34}$ in the odd manifold. Degeneracy therefore occurs for $\varepsilon_{12}=-\varepsilon_{34}$ (along the antidiagonal) and $\varepsilon_{12}=\varepsilon_{34}$ (along the diagonal), respectively. Away from these conditions, the finite longitudinal term tilts the rotation axis, reduces contrast, increases the return probability and frequency, and enhances charge-noise sensitivity.

Taken together, these measurements show that coherent oscillations can be observed even when the two chains are not individually degenerate, provided the relevant global parity manifold remains degenerate. Measuring both manifolds is therefore necessary to identify the simultaneous Majorana sweet spot of the two chains. At the sweet spot, the matching Rabi frequencies further show that the effective inter-chain coupling is approximately parity independent, consistent with a small overlap between Majorana wavefunctions (see Methods~\ref{methods:consequences_imperfect_majoranas}). 

\section*{Robustness of parity oscillations}\label{detunings}
\begin{figure*}[h!]
    \centering
    \includegraphics[width=1\textwidth]{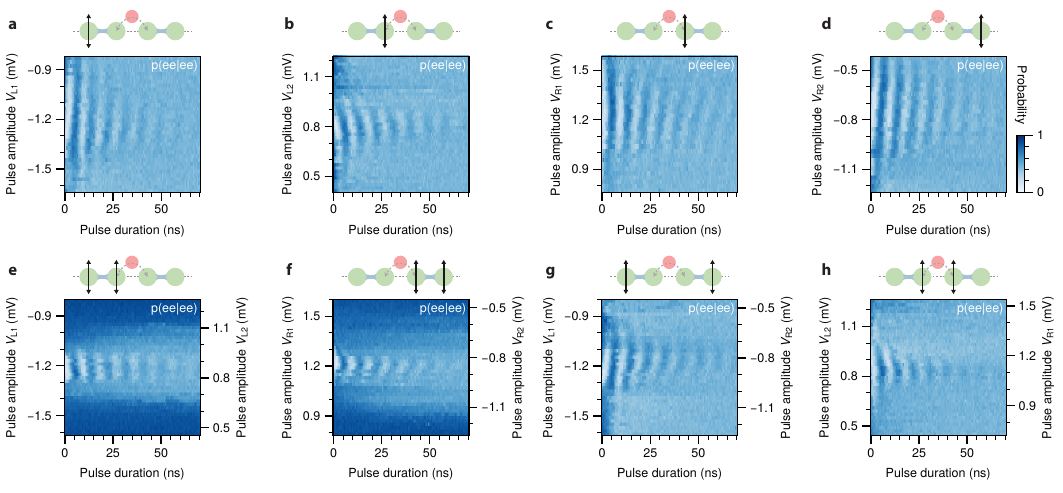}
    \caption{\textbf{Robustness of coherent parity oscillations under detuning.}
    $\peeee$ as a function of pulse duration for different detuning configurations, as indicated by the schematics. The coupler pulse amplitude is fixed at $\VC = \SI{9.8}{mV}$.
    \textbf{a-d,} Detuning individual quantum dots. Oscillations remain visible over a broader range of gate voltages compared to panels \textbf{e-h}.
    \textbf{e,f,} Detuning the left (\textbf{e}) and right (\textbf{f}) Kitaev chains. Increasing chain detuning suppresses the oscillations and stabilizes the initial state.
    \textbf{g,h,} Simultaneous detuning of the outer (\textbf{g}) and inner (\textbf{h}) quantum dots of both chains, resulting in a more rapid reduction of oscillation visibility.
    }
    \label{fig:4}
\end{figure*}

Finally, we probe how the coherent dynamics responds to detuning of the chain QDs (Fig.~\ref{fig:4}). At the sweet spot in a single two-site chain, the even--odd degeneracy is protected against detuning of either QD individually; a splitting is generated only when both QDs of the same chain are detuned~\cite{leijnse_parity_2012,dvir_realization_2023}. This provides a direct way to test the partial protection of minimal Kitaev chains.

Detuning one QD at a time gradually reduces oscillation visibility, with the signal decaying toward $\peeee\approx0.5$ (Fig.~\ref{fig:4}a--d), consistent with increased dephasing and no sizable longitudinal splitting. Detuning both QDs of the same chain suppresses oscillations more rapidly and increases the return probability toward unity (Fig.~\ref{fig:4}e,f), as expected for a finite intra-chain splitting that tilts the rotation axis toward $Z$. Detuning one QD in each chain suppresses oscillations on a similar scale but keeps the decayed signal near $\peeee\approx0.5$, consistent with reduced coherence while individual-chain degeneracies remain preserved (Fig.~\ref{fig:4}g,h). Additional measurements are shown in Extended Data, including results for initialization in $\ket{oo}$ (Fig.~\ref{fig:ed_detunings}) and coherent parity oscillations measured in a different tuning configuration (Fig.~\ref{fig:ed_reproduction}).

These measurements show that coherent oscillations require all chain QDs near resonance, confirming that the encoded qubit is distributed across every site of the two chains.
They also highlight the role of the partial protection provided by poor man's Majoranas. Detuning a single QD away from resonance (Fig.~\ref{fig:4}a--d) removes the sweet spot protection and suppresses the oscillations. Detuning one QD in each chain (Fig.~\ref{fig:4}g,h) reduces the coupled chains to an effective double QD and provides a direct comparison to a conventional charge qubit~\cite{hayashi_coherent_2003,petersson_quantum_2010}, for which no coherent oscillations are resolved under the same measurement conditions.

\section*{Conclusion}\label{Disc}
We have demonstrated a Majorana qubit by measuring coherent Rabi oscillations between parity states of two coupled minimal Kitaev chains. At the sweet spot, the oscillations in both global parity manifolds occur at matching frequencies, as predicted for a qubit encoded in isolated Majorana zero modes~\cite{tsintzis_majorana_2024}.
We show that the qubit frequency and coherence depend on inter-chain coupling and quantum-dot detunings in agreement with the specific partial protection mechanism of Majorana modes in minimal Kitaev chains~\cite{tsintzis_majorana_2024,pan_rabi_2025}. Our theoretical modeling assuming quasistatic charge noise reproduces the main observations (Fig.~\ref{fig:ed_theory_even}, Fig.~\ref{fig:ed_theory_odd}).

Although the protection in two-site chains is limited, this platform provides a route to longer Kitaev chains, where the sensitivity to perturbations in the idle state is predicted to decrease exponentially with the number of sites~\cite{kitaev_unpaired_2001,bordin_enhanced_2025,ten_haaf_observation_2025,knapp_dephasing_2018,mishmash_dephasing_2020}. 
We note that operations relying on coupling Majorana modes that lift the qubit degeneracy are not topologically protected, independent of chain length. This is a limitation of the control operation, not of the encoding itself: our encoding can be used for a topologically protected Majorana qubit in longer chains. Non-topological operations nevertheless remain useful in Majorana-based quantum computation, for example for preparing non-Clifford resources such as magic states~\cite{sarma_majorana_2015}.
Future experiments with longer chains will enable direct tests of the predicted protection scaling~\cite{kitaev_unpaired_2001,sau_realizing_2012,tsintzis_majorana_2024,pan_rabi_2025}, as well as fusion and braiding protocols aimed at demonstrating non-Abelian statistics~\cite{ivanov_non-abelian_2001,plugge_majorana_2017,karzig_scalable_2017, boross_braiding-based_2024, tsintzis_majorana_2024,miles_braiding_2026}.

\section*{Data Availability and Code Availability}
All raw data measured on the device presented in this work at the time of writing, the data processing and plotting code, and the code used for the theory
calculations are available at
\url{https://zenodo.org/records/21280458}.

\section*{Acknowledgements}
We thank S. Bhatti, D. Prete, S. Goswami, Y. Zhang, W. Huisman, S. Roelofs, A. Chatterjee, A. Akhmerov, G. Wang, G.P. Mazur, K. Flensberg, and G.O. Steffensen for useful discussions, O.W.B. Benningshof, J.D. Mensingh, T. Orton, and E. van der Wiel for technical assistance with the cryogenic setup, R. Schouten and R. Birnholtz for technical assistance with the measurement electronics, M. Fischer and B. van Asten for technical assistance with the aluminum evaporator, S. Gazibegovic for contributing to nanowire growth.
This work has been supported by the Dutch Organization for Scientific Research (NWO) and Microsoft Corporation Station Q. 

\section*{Author contributions}
The sample was fabricated by B.R. and N.v.L.
Measurements were performed by F.Z. and B.R. under supervision of N.v.L.
The experiment was designed by F.Z., B.R., and N.v.L.
The data were analyzed by F.Z., B.R., N.v.L., and A.L.
The experimental setup was designed and implemented by F.Z., B.R., and N.v.L.
All authors from Delft University of Technology contributed to the understanding and interpretation of the experimental results through regular discussions.
The manuscript was prepared by F.Z., B.R., N.v.L., and L.P.K. with input from all authors.
Modeling and simulations of the system were done by J.D.T.L., S.M., and M.W. with the help of A.L. and V.P.M.S., under the supervision of M.W. Theory sections of the Methods were written by M.W.
The InSb nanowire growth was performed by G.B. and E.P.A.M.B.
The project was supervised by L.P.K. 

\section*{Competing interests}
The authors declare no competing interests.
\clearpage

\section*{Extended Data}

\setcounter{figure}{0}
\renewcommand{\thefigure}{ED\arabic{figure}}
\renewcommand{\theHfigure}{ED\arabic{figure}}
\begin{figure*}[ht!]
    \centering
    \includegraphics[width=1\textwidth]{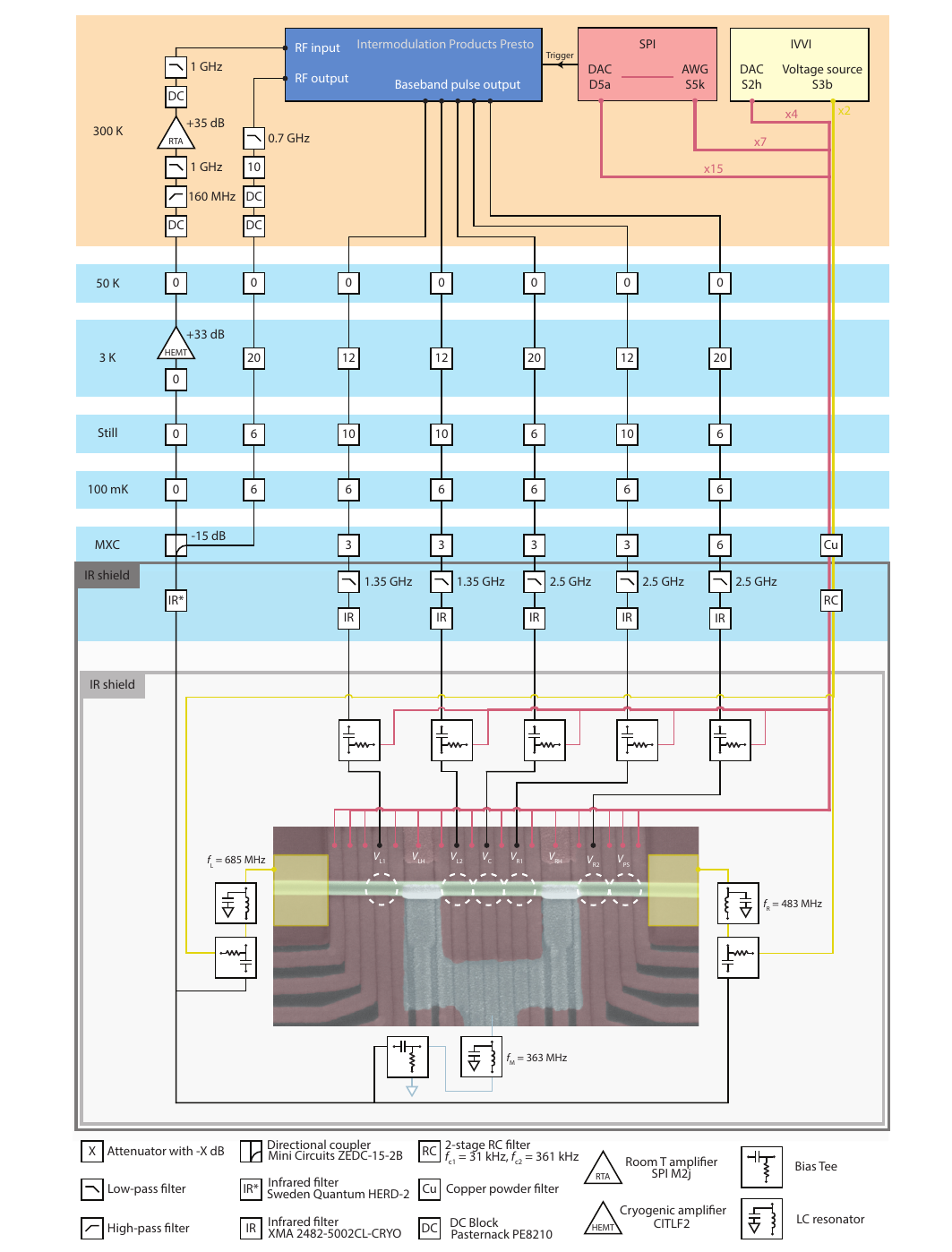}
    \caption{\textbf{Measurement setup and wiring diagram.} Wiring diagram for the DC and RF lines used to measure the device, with properties and components detailed in the legend. Additional details can be found in the Methods section.
    \label{fig:ed_measurementsetup}
    }
\end{figure*}

\begin{figure*}[ht!]
    \centering
    \includegraphics[width=1\textwidth]{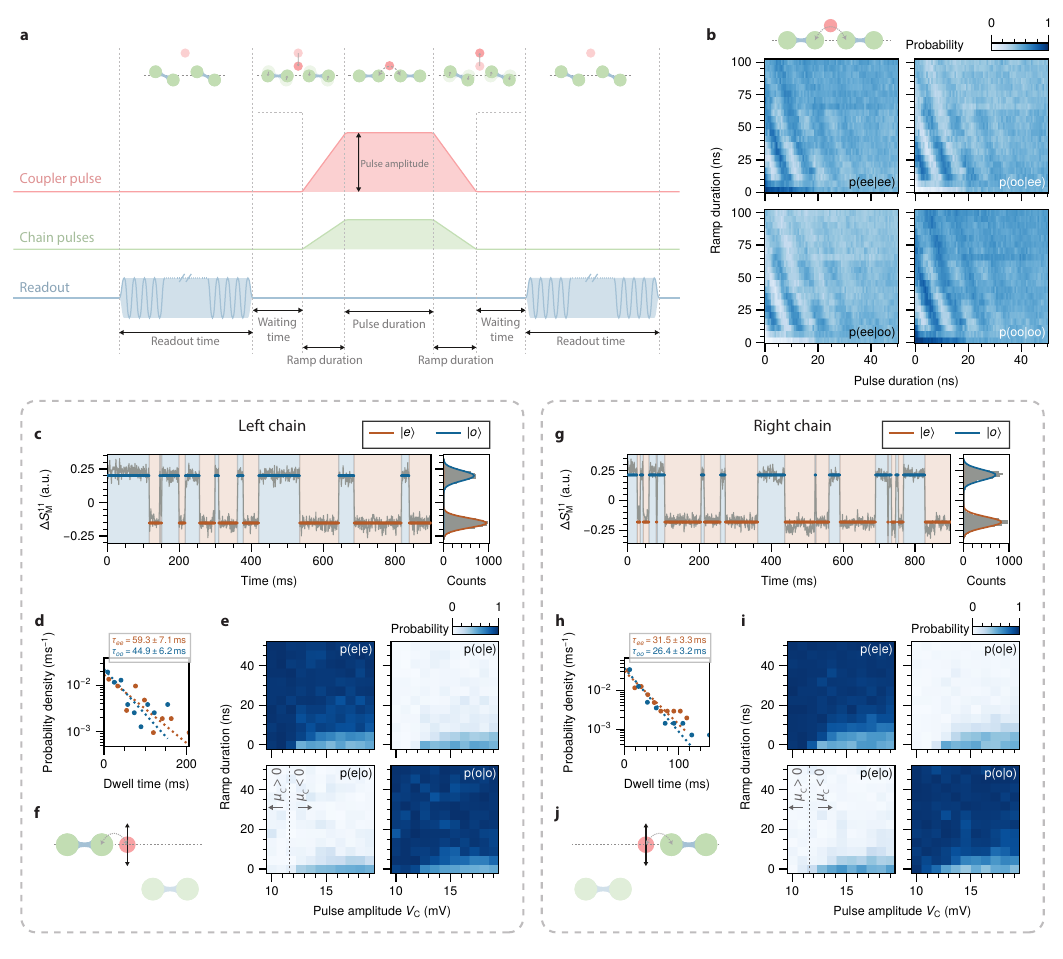}
    \caption{\textbf{Varying ramp time.}
    \textbf{a,} Detailed schematic of the pulse sequence, showing the various stages of the protocol.
    \textbf{b,} $\peeee$, $\pooee$, $\peeoo$, and $\poooo$ as a function of pulse duration and pulse ramp time. Parity oscillations remain visible even for slow ramp duration up to approximately $\SI{75}{\nano s}$. An oscillation as a function of ramp duration is also visible at zero pulse duration, indicating that the qubit evolves partially during the ramp of the coupling pulse. 
    \textbf{c,} Time trace and corresponding histogram of the readout signal $\Delta\SM$ with the left chain on resonance and the right-chain QDs off-resonance, using \SI{400}{\micro\second} readout time. Transitions between $\ket{e}$ and $\ket{o}$ are attributed to quasiparticle poisoning. 
    \textbf{d,} Dwell time histogram from \textbf{c}. 
    \textbf{e,} $\pee$, $\poe$, $\peo$, and $\poo$ while varying the pulse amplitude $\VC$ and ramp time. For fast ramps ($\lesssim \SI{5}{ns}$), parity mixing is observed between the chain and the dot. We attribute this to a diabatic excitation of the coupled system, which rapidly decoheres before the final measurement. In the main text, we use ramp times of \SI{25}{ns} to avoid these excitations. For pulse amplitudes $\lesssim \SI{11.5}{mV}$, no parity mixing is observed, which we use to roughly identify when QDC passes through resonance. 
    \textbf{f,} Schematic of the configuration for the measurements in \textbf{c-e}. 
    \textbf{g-j,} Same as \textbf{c-f}, but for the right chain, with the QDs of the left chain off-resonance. In panel \textbf{i,} parity mixing between the chain and QDC is observed for ramp times $\lesssim \SI{10}{ns}$, indicating a slightly weaker coupling between QDC and the right chain versus the left chain.
}
    \label{fig:ed_rampduration}
\end{figure*}

\begin{figure*}[ht!]
    \centering
    \includegraphics[width=1\textwidth]{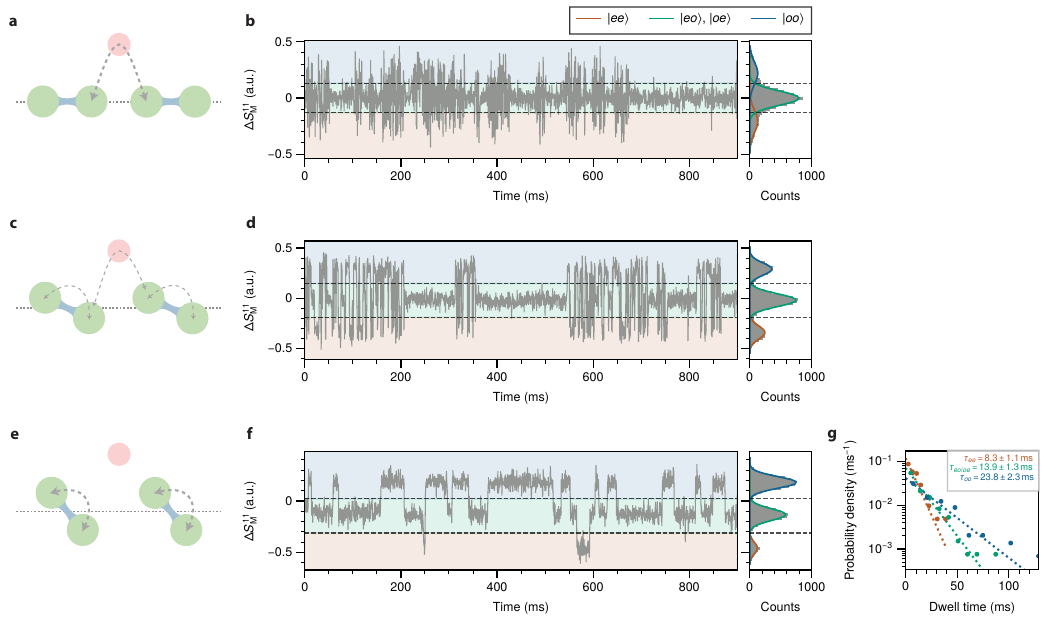}
    \caption{\textbf{Readout characterization.} 
    Readout characterization using time traces of the reflected resonator signal $\Delta\SM$, measured for three different amounts of detuning with the coupler QD kept fixed far from resonance. Thresholds used for single-shot state identification are obtained by fitting a Gaussian mixture model. The detuning configuration in (\textbf{e-g}) is used for qubit readout.
    \textbf{a,} Schematic representation of the readout configuration with both chains on resonance.
    \textbf{b,} Time trace of $\Delta\SM$ and corresponding histogram for the readout configuration in \textbf{a}. Residual coupling between the chains results in inter-chain hopping that is faster than the readout time of $\SI{500}{\micro s}$, preventing readout in a single shot. The two quantum capacitance signals corresponding to the $\ket{ee}$ and $\ket{oo}$ states are partially averaged which makes them difficult to distinguish from the $\ket{oe}$ and $\ket{eo}$ states.
    \textbf{c,} As in \textbf{a}, but with both chains detuned from resonance.
    \textbf{d,} Time trace of $\Delta\SM$ and corresponding histogram for the readout configuration in \textbf{c}. The slight detuning suppresses inter-chain hopping, though it remains too fast for high-fidelity single-shot readout. Two distinctive regions can be observed: When the system is in the global even state, rapid jumps between $\ket{oo}$ and $\ket{ee}$ are visible. When the system is poisoned, it enters the global odd manifold. Transitions between $\ket{oe}$ and $\ket{eo}$ are not resolved because they produce similar readout signals.
    \textbf{e,} As in \textbf{c} but with both chains detuned further from resonance.
    \textbf{f,} Time trace of $\Delta\SM$ and corresponding histogram for the readout configuration in \textbf{e}. The two computational states $\ket{ee}$ and $\ket{oo}$ correspond to the red and blue signals, and can now clearly be resolved. The green signal contains both leakage states of the odd global parity manifold ($\ket{eo}$ and $\ket{oe}$). Due to the antisymmetric detuning where the ground state of one dot is full and the other is empty, the $\ket{oo}$ state is energetically favored over the $\ket{ee}$ state, resulting in a skewed population between the two. Transitions are mainly observed between the even and odd global manifolds.
    \textbf{g,} Dwell time histogram of the three observed states, indicating the difference in lifetimes. The lifetime of each state substantially exceeds the $\SI{500}{\micro s}$ readout duration, enabling reliable single-shot state discrimination.
}
    \label{fig:ed_readout}
\end{figure*}

\begin{figure*}[ht!]
    \centering
    \includegraphics[width=1\textwidth]{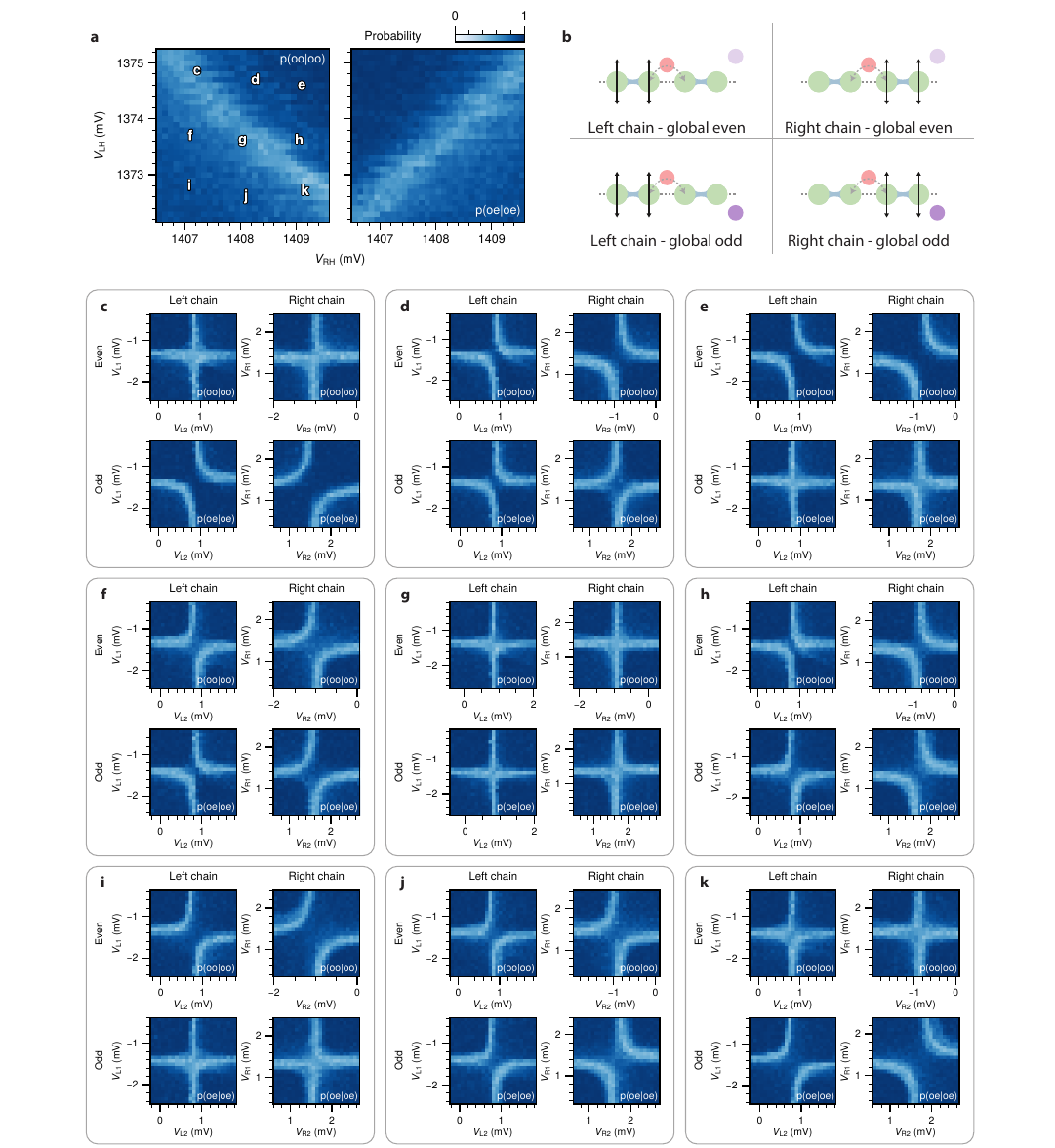}
    \caption{\textbf{Pulsed charge-stability diagrams across parameter space.}
    \textbf{a,} Map of $\poooo$ (left) and $\poeoe$ (right) as a function of the left and right hybrid-gate voltages ($\VLH$ and $\VRH$). The diagonal and antidiagonal features of enhanced parity mixing correspond to degeneracies between the qubit basis states. The points labeled c–k indicate the $\VLH$,$\VRH$ values used in the remaining panels.
    \textbf{b,} Schematic representation of the four configurations measured at each point: left- and right-chain pulsed charge-stability diagrams in the global even and odd parity manifolds.
    \textbf{c-k,} Pulsed charge-stability diagrams measured for the left and right chains in the even (top) and odd (bottom) parity manifolds. Unlike conventional charge-stability diagrams referenced to the Fermi energy in a normal lead, here each chain provides the energy reference for the other. As such, the measured features reflect the difference between the excitation energies of the two individual chains. Crossings are therefore observed simultaneously for the left and right chains when the splittings of the two chains compensate each other, corresponding to the qubit degeneracy condition. Consequently, the appearance of a crossing in a single parity manifold does not imply that the corresponding chain is tuned to its Majorana sweet spot. Indeed, crossing-like features are observed in only one manifold in panels c, e, i and k. The orientation of the avoided crossings depends on the parity manifold: in the even manifold, the left and right chains exhibit the same orientation, whereas in the odd manifold the orientation is reversed, reflecting the sign difference $\varepsilon_{12} \pm \varepsilon_{34}$ between the two manifolds. Only near the Majorana sweet spot (panel g) are crossings observed for both chains in both parity manifolds simultaneously, providing a unique signature that both chains are individually tuned to their sweet spots.
    }
    \label{fig:ed_PCSDs}
\end{figure*}

\begin{figure*}[ht!]
    \centering
    \includegraphics[width=1\textwidth]{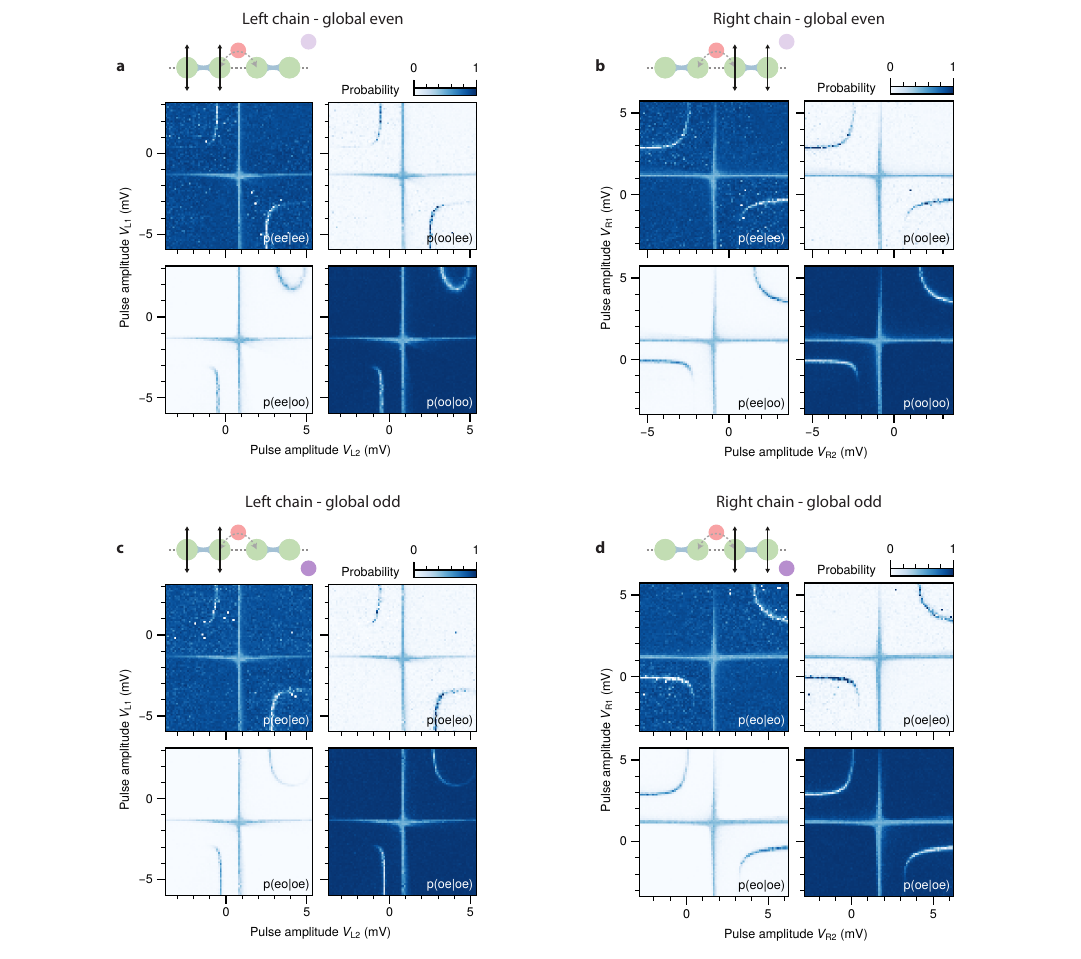}
    \caption{\textbf{Pulsed charge-stability diagrams at the Majorana sweet spot.}
    Conditional probabilities at the simultaneous Majorana sweet spot for the two chains in both the even and odd global parity manifolds, corresponding to Fig.~\ref{fig:ed_PCSDs}g but extended to larger pulse amplitudes. Additional outer lines appear when the energy of the detuned chain is aligned with the first excited state of the other chain. The quadrant in which they appear depends on the initial state of the system, as well as the global parity manifold (see Methods~\ref{Methods_pulsedCSD}).
    \textbf{a,} Left chain in the even manifold.
    \textbf{b,} Right chain in the even manifold.
    \textbf{c,} Left chain in the odd manifold.
    \textbf{d,} Right chain in the odd manifold.
    }
    \label{fig:ed_PCSDs_excited}
\end{figure*}

\begin{figure*}[ht!]
    \centering
    \includegraphics[width=1\textwidth]{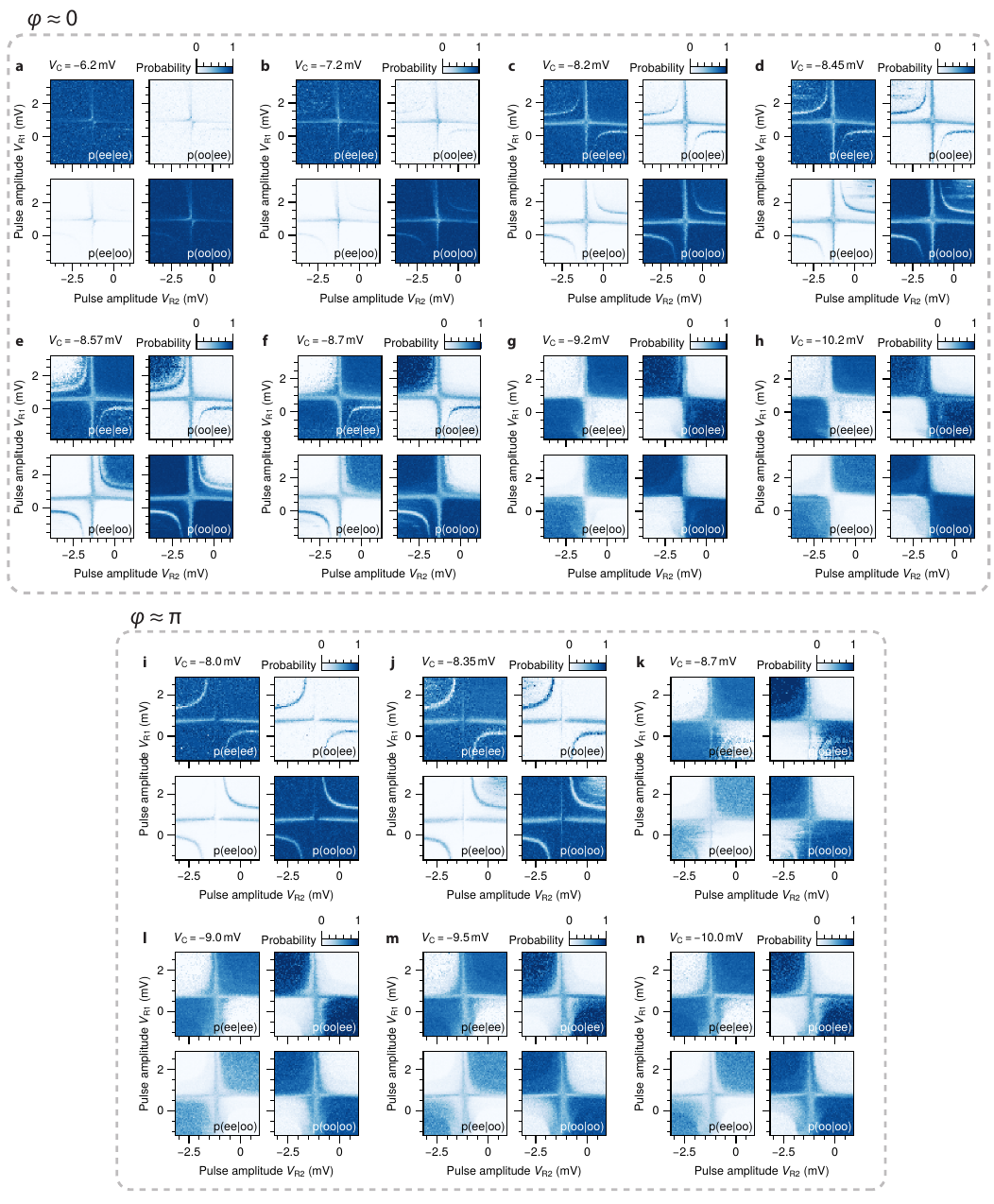}
    \caption{\textbf{Pulsed CSDs varying pulse amplitude on \QDC.}
    Pulsed charge-stability diagrams showing $\peeee$, $\pooee$, $\peeoo$, and $\poooo$ as a function of the right-chain gate voltages ($\VRL,\VRR$), for various values of the coupler QD pulse amplitude $\VC$ for a fixed pulse duration of \SI{10}{ns}. The value of $\VC$ is indicated above each panel. The device configuration is the same as in Fig.~\ref{fig:ed_reproduction}.
    \textbf{a-h,} The superconducting phase difference between the chains is set to $\varphi \approx 0$. As $\VC$ is made more negative, the coupler QD is first brought closer to resonance and parity mixing along the degeneracy lines increases. Eventually, when the coupler QD is near or below resonance (around $\VC = \SI{-9}{mV}$), the final state is dictated by the ground state of the chain in the relevant sector of the charge-stability diagram irrespective of the initial state. We note that the chain appears to move from CAR-like at low pulse amplitudes ($\tR < \DeltaR$, \textbf{a}) across a sweet spot ($\tR \approx \DeltaR$, \textbf{b-e}) to ECT-like at high pulse amplitudes ($\tR > \DeltaR$, \textbf{f-h}), which we attribute to uncompensated cross-capacitance between $\VC$ and $\VRH$.
    \textbf{i-n,} The superconducting phase difference between the chains is set to $\varphi \approx \pi$. The main difference is that the visibility of \QDRR is suppressed, as predicted by theory (Methods~\ref{methods:consequences_imperfect_majoranas}).
}
    \label{fig:ed_pulsedCSDsQDC}
\end{figure*}

\begin{figure*}[ht!]
    \centering
    \includegraphics[width=1\textwidth]{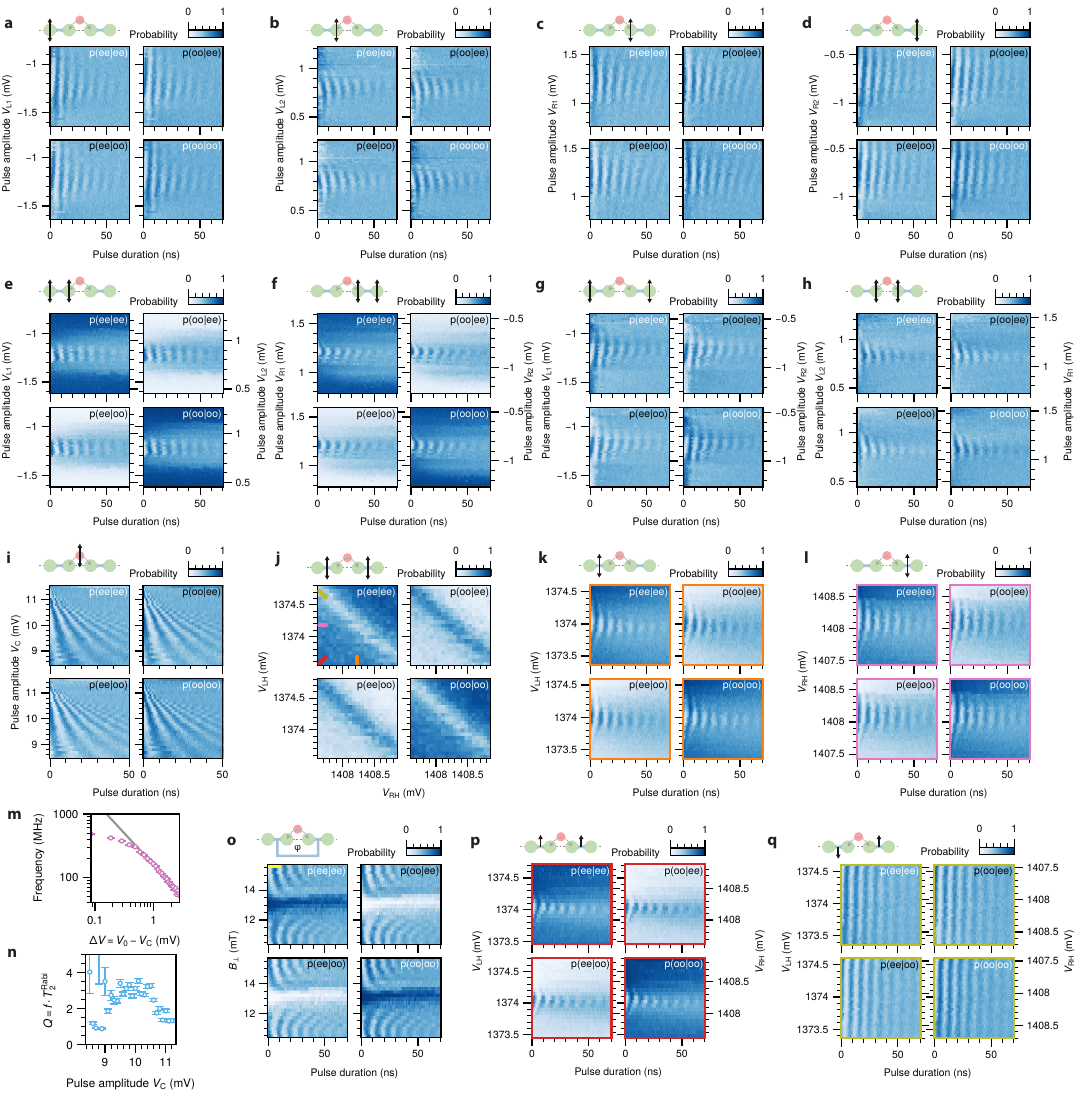}
    \caption{\textbf{Full set of probability matrices for the even manifold.}
    $\peeee$, $\pooee$, $\peeoo$, and $\poooo$ as a function of pulse duration for various detunings, as indicated by the schematics, corresponding to measurements shown in the main text. Unless stated otherwise, measurements are taken at $\VC = \SI{9.8}{mV}$ and perpendicular field $B_\perp = \SI{15.6}{mT}$ (yellow mark in panel o), corresponding to a phase difference $\varphi \approx 0$ between the chains.
    \textbf{a-d,} Detuning individual quantum dots. 
    \textbf{e,f,} Detuning the left- (\textbf{e}) and right-chain QDs (\textbf{f}).
    \textbf{g,h,} Simultaneous detuning of the outer (\textbf{g}) and inner (\textbf{h}) QDs of both chains.
    \textbf{i,} Varying coupling pulse duration and amplitude.
    \textbf{j,} Varying $\VLH$ and $\VRH$ for a fixed pulse duration of \SI{18}{ns}. The colored marks indicate sweep directions for panels \textbf{k,l,n,o}.
    \textbf{k,l,} Detuning the left (\textbf{k}) and right (\textbf{l}) chain via $\VLH$ and $\VRH$.
    \textbf{m,} Oscillation frequency as a function of pulse amplitude in logarithmic scale, where $V_\mathrm{0}$ is the gate voltage at which \QDC is on resonance. The gray line fits $f = K/(V_\mathrm{0}-\VC)$ with $K = \SI{156.6 \pm 0.3}{\MHz\cdot\mV}$ and $V_\mathrm{0} = \SI{11.285 \pm 0.002}{mV}$ for $\VC < \SI{10.8}{mV}$.
    \textbf{n,} Quality factor of the oscillations as a function of $\VC$.
    \textbf{o,} Dependence on the out-of-plane magnetic field $B_\perp$.
    \textbf{p,q,} Symmetric (\textbf{p}) and antisymmetric (\textbf{q}) detuning of $\VLH$ together with $\VRH$.
}
    \label{fig:ed_detunings}
\end{figure*}

\begin{figure*}[ht!]
    \centering
    \includegraphics[width=1\textwidth]{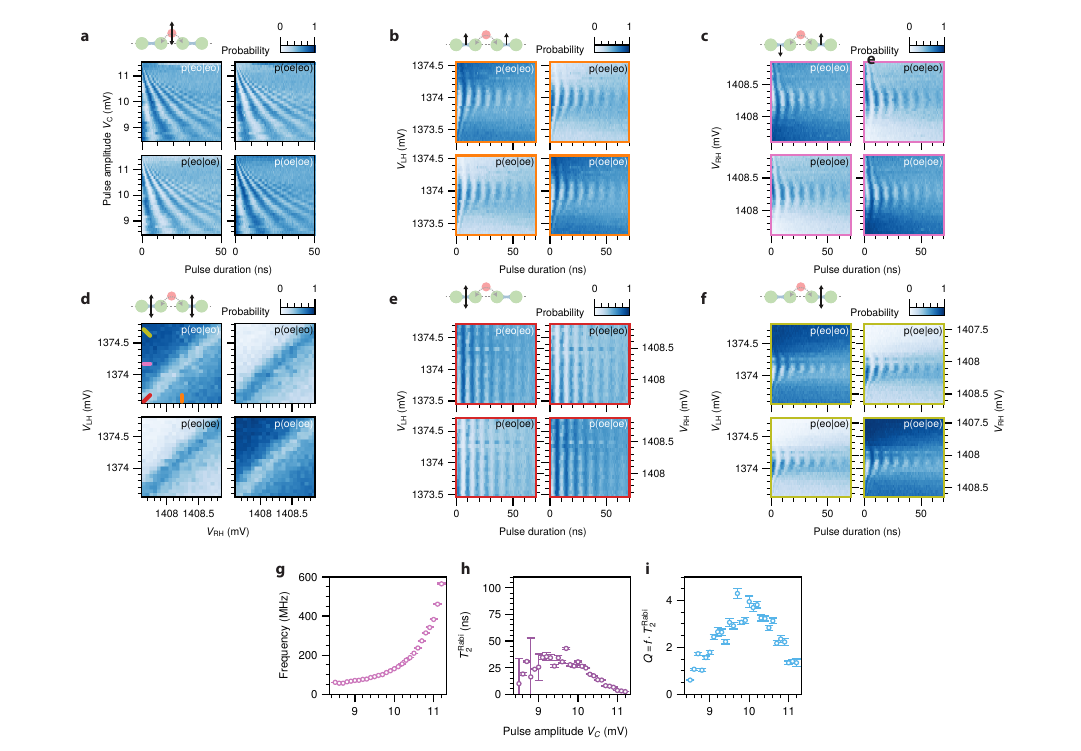}
    \caption{\textbf{Full set of probability matrices for the odd manifold.}
    $\peoeo$, $\poeoe$, $\peooe$, and $\poeeo$  as a function of pulse duration for various detunings, as indicated by the schematics, corresponding to measurements shown in the main text.
    \textbf{a,} Varying coupling pulse duration and amplitude $\VC$.
    \textbf{b,c,} Detuning the left (\textbf{b}) and right (\textbf{c}) chain via $\VLH$ and $\VRH$.
    \textbf{d,} Varying $\VLH$ and $\VRH$ for a fixed pulse duration of \SI{18}{ns}. The colored marks indicate sweep directions for panels \textbf{b,c,e,f}.
    \textbf{e,f,} Symmetric (\textbf{e}) and antisymmetric (\textbf{f}) detuning of $\VLH$ together with $\VRH$.
    \textbf{g-i,} Oscillation frequency (\textbf{g}), Rabi decay time $T_2^{\mathrm{Rabi}}$ (\textbf{h}) and quality factor $Q$ (\textbf{i}) extracted from panel \textbf{a} as a function of $\VC$.
}
    \label{fig:ed_detunings_odd}
\end{figure*}

\begin{figure*}[ht!]
    \centering
    \includegraphics[width=1\textwidth]{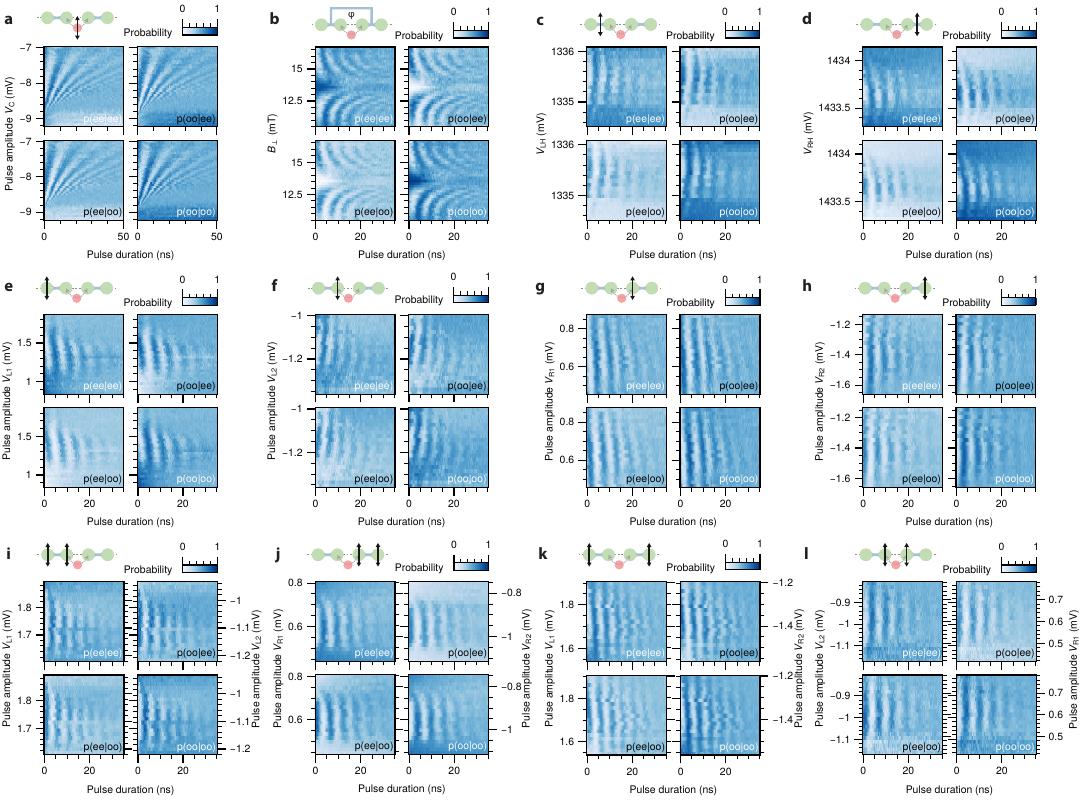}
    \caption{\textbf{Parity oscillations in the even manifold for different QD orbitals and Andreev bound states.}
    $\peeee$, $\pooee$, $\peeoo$, and $\poooo$ as a function of pulse duration for various detunings, as indicated by the schematics. Note that for this dataset, the chains were tuned using pulsed charge-stability diagrams but were not compared to the odd manifold. As such, it is unclear if the even and odd states of each chain are degenerate.  
    \textbf{a,} Varying coupling pulse duration and amplitude $\VC$. In contrast to the main text, \QDC starts detuned below the Fermi level instead of above but is likewise pulsed closer to resonance for qubit evolution.
    \textbf{b,} Dependence on the out-of-plane magnetic field $B_\perp$. In this case, we observe decoherence around $\varphi = \pi$ ($B_\perp \approx \SI{13.5}{mT}$) which we attribute to imperfect tuning.
    \textbf{c,d,} Detuning the left (\textbf{c}) and right (\textbf{d}) chain via $\VLH$ and $\VRH$.
    \textbf{e-h,} Single-dot detuning of $\VLL$, $\VLR$, $\VRL$ and $\VRR$.
    \textbf{i,j,} Detuning the left- (\textbf{i}) and right-chain QDs (\textbf{j}).
    \textbf{k,l,} Simultaneous detuning of the outer (\textbf{k}) and inner (\textbf{l}) quantum dots of both chains.
}
    \label{fig:ed_reproduction}
\end{figure*}

\newpage
\begin{figure}
    \centering
    \includegraphics[width=\linewidth]{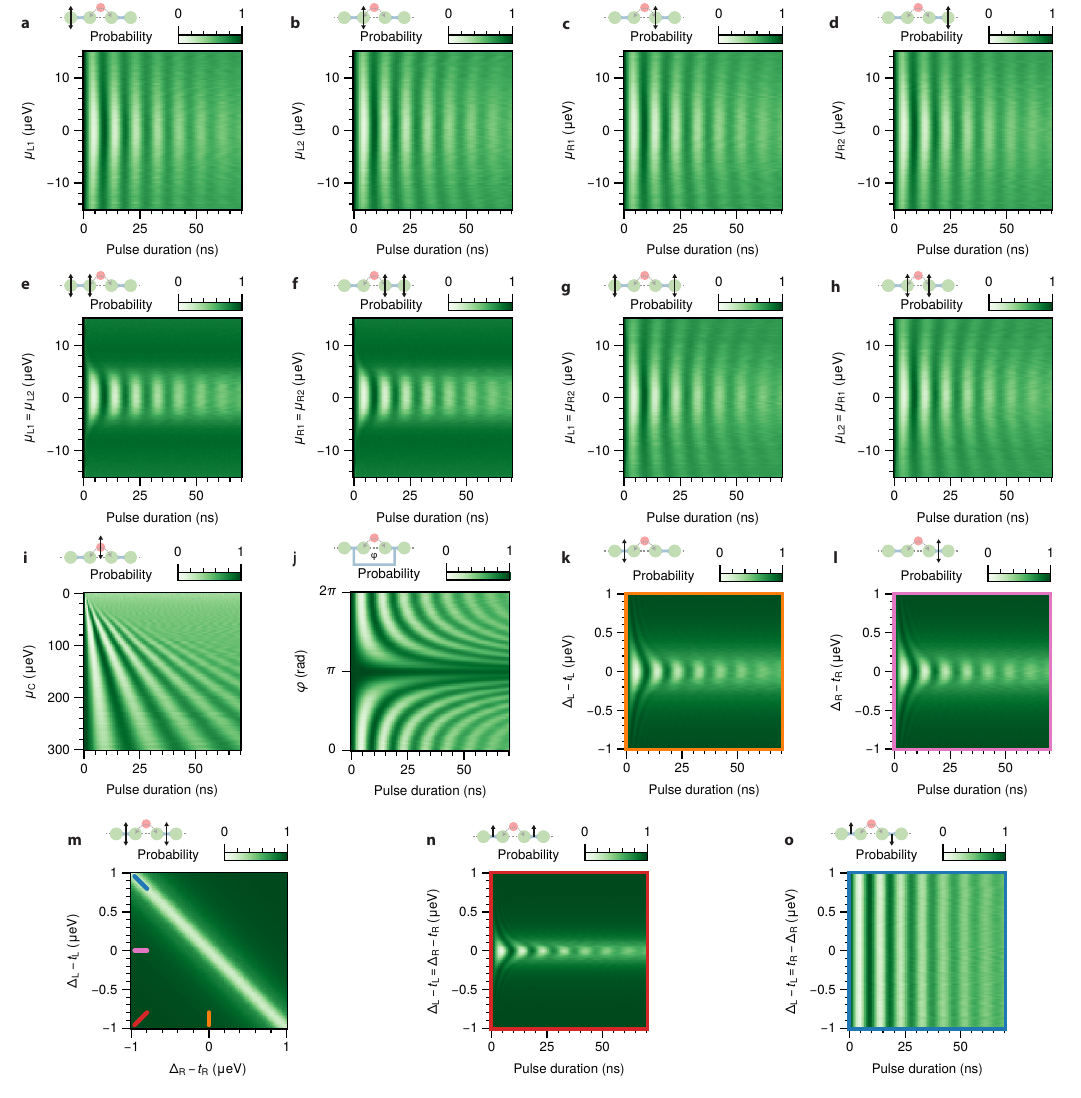}
    \caption{\textbf{Numerical simulations of noise-averaged return probability for the even sector of two coupled minimal Kitaev chains.} The system is initialized in the even state $|ee\rangle$ and evolved for a square-pulse duration $T$ under the Hamiltonian in Eq.~\eqref{model:pmm_hamiltonian} with the parameter indicated on each vertical axis held at its pulsed value.
    Unless otherwise specified, the middle-dot chemical potential is fixed at $\muC^*=130\,\mu\mathrm{eV}$. All parameters are given in Table~\ref{tab:detuning_simulation_parameters}. Each pixel is averaged over $N=1000$ realizations of quasistatic Gaussian noise.
    \textbf{a-d,} Detuning individual chemical potentials $\muLL$, $\muLR$, $\muRL$, and $\muRR$.
    \textbf{e,f,} Detuning of the left (\textbf{e}) and right (\textbf{f}) chains via $\muLL=\muLR$ and $\muRL=\muRR$.
    \textbf{g,h,} Detuning of the outer (\textbf{g}) and inner (\textbf{h}) chemical-potential pairs via $\muLL=\muRR$ and $\muLR=\muRL$.
    \textbf{i,} Rabi oscillations obtained by varying $\muC$ and keeping the two chains at the sweet spot.
    \textbf{j,} Dependence on the superconducting phase $\varphi$ at fixed $\muC^*$ at the sweet spot.
    \textbf{k,l,} Varying the CAR and ECT amplitudes in the left (\textbf{k}) and right (\textbf{l}) chains.
    \textbf{m,} Return probability after a $\pi$ pulse as a function of the CAR and ECT interactions in both chains.
    \textbf{n,o,} Symmetric (\textbf{n}) and antisymmetric (\textbf{o}) detuning of the CAR and ECT amplitudes in both chains.
}
    \label{fig:ed_theory_even}
\end{figure}

\begin{figure}
    \centering
    \includegraphics[width=\linewidth]{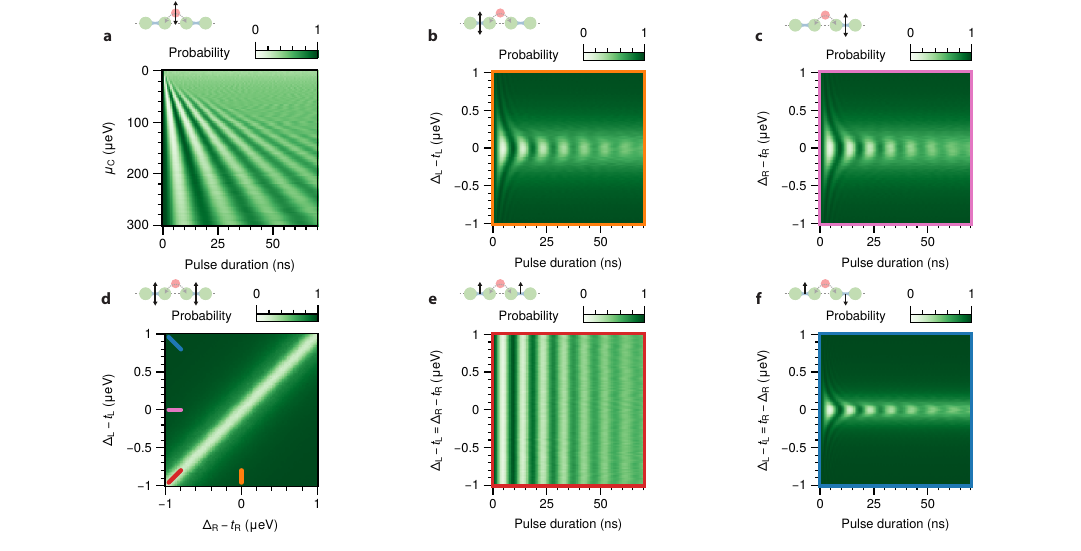}
    \caption{\textbf{Numerical simulations of noise-averaged return probability for the odd sector of two coupled minimal Kitaev chains.} All parameters are given in Table \ref{tab:detuning_simulation_parameters}.
    Each pixel is averaged over $N=1000$ realizations of quasistatic Gaussian noise.
    \textbf{a,} Rabi oscillations obtained by varying $\muC$ and keeping the two chains at the sweet spot.
    \textbf{b,c,} Varying the CAR and ECT amplitudes in the left (\textbf{b}) and right (\textbf{c}) chains.
    \textbf{d,} Return probability after a $\pi$ pulse as a function of the CAR and ECT interactions in both chains.
    \textbf{e,f,} Symmetric (\textbf{e}) and antisymmetric (\textbf{f}) detuning of the CAR and ECT amplitudes in both chains.
}
    \label{fig:ed_theory_odd}
\end{figure}

\begin{figure}
    \includegraphics[width=\linewidth]{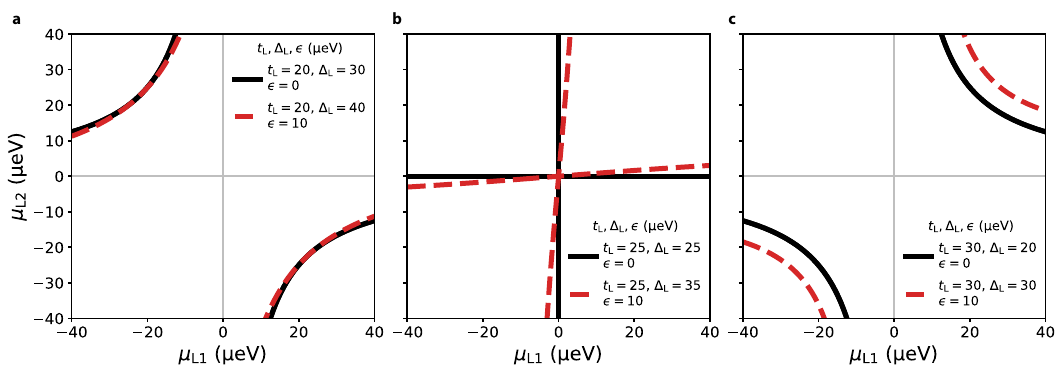}
    \caption{Lines of zero longitudinal splitting in the $\muLL$-$\muLR$-plane from Eq.~\eqref{eq:analytical_csd} for one parity sector. $\epsilon$ is a fixed energy splitting of the right chain. Parameters are given in the legends. Both parameter sets with $\epsilon=0$ and $\epsilon\neq 0$ show qualitatively similar behavior, but only the case $\epsilon=0$ corresponds to a true CAR to ECT transition in the left chain.}\label{fig:tuning_theory}
\end{figure}

\clearpage
\section*{Methods}\label{Methods}
\renewcommand{\thesection}{}
\renewcommand{\thesubsection}{\Roman{subsection}}
\renewcommand{\thesubsubsection}{\Alph{subsubsection}}

\let \savenumberline \numberline
\def \numberline#1{\savenumberline{#1.}}
\subsection{Device fabrication}\label{Methods_fab}
The device is fabricated using a shadow-wall lithography process described previously~\cite{heedt_shadow-wall_2021,mazur_spin-mixing_2022,van_loo_single-shot_2026}, with key parameters summarized here. 

The base substrate consists of high-resistivity Si with a \SI{100}{\nano m} Si$_3$N$_4$ layer grown by low-pressure chemical vapor deposition. Gate electrodes are defined by electron-beam lithography (EBL) followed by metal evaporation ($3/12$\,nm Ti/Pd bilayer), and connected to \SI{50}{\nano m} W bond pads fabricated by EBL and RF sputtering. The gates are covered by a $20/10$\,nm Al$_2$O$_3$/HfO$_2$ dielectric deposited by atomic layer deposition. 

An InSb nanowire with a diameter of approximately \SI{85}{\nano\meter} (growth details in Ref.~\cite{badawy_high_2019}) is placed on top of the dielectric using a nanomanipulator. Prior to superconductor deposition, the nanowire oxides are removed in situ using hydrogen radicals. A 15\,nm Al film is then deposited at a substrate temperature of 138\,K under a $30^\circ$ angle with respect to the substrate using metal evaporation, with device patterning defined by an HSQ shadow mask. Next, a 10\,nm AlO$_x$ capping layer is evaporated at the same temperature and angle. The residual Al film outside the device area is subsequently removed via a Transene type D wet etch~\cite{van_loo_single-shot_2026}. Finally, metal contacts are defined by EBL, followed by Ar milling for surface treatment, and metal evaporation ($10/90$\,nm Cr/Au) to contact the nanowire for transport measurements.

\subsection{Measurement setup}\label{Methods_setup}
Fig.~\ref{fig:ed_measurementsetup} shows a schematic of the room-temperature electronics and cryogenic wiring used for control and readout. The system is divided into temperature stages (300\,K, 50\,K, 3\,K, still, 100\,mK and mixing chamber (MXC)). At room temperature, DC voltages for gate control are generated by digital-to-analog converters in the SPI and IVVI racks. An arbitrary waveform generator built in-house (S5k module for SPI rack) is used to generate sawtooth signals for fast gate sweeps during tune-up.
For qubit readout, a Presto unit from Intermodulation Products is used to generate and demodulate RF signals. Two RF outputs of the Presto unit are combined using a splitter, where one is used for readout while the other provides an off-resonant tone at $\SI{580}{\mega\hertz}$ to thermalize the readout chain (see Methods~\ref{Methods_pulses}). Additionally, the Presto creates arbitrary waveforms with a sampling rate of \SI{500}{\mega Sa\per\second} for fast control of the quantum dot chemical potentials.

Along the cryostat, each RF line is progressively thermalized and filtered using a combination of attenuators, low-pass, high-pass and infrared (IR) filters, while DC lines are filtered using RC and copper powder filters at low temperatures. Several layers of shielding are used to reduce noise and protect the sample from radiation that could generate non-equilibrium quasiparticles in the superconducting film. Attenuation and cutoff frequencies are indicated in Fig.~\ref{fig:ed_measurementsetup}. Details of the other components are specified in the legend. 
The RF and DC lines are combined via bias tees on the PCB (\SI{72}{\kilo\hertz} cutoff frequency) and connected to the quantum dot plunger gates. The RF input and output lines are coupled via a directional coupler and connected to a separate frequency-multiplexing chip. Three spiral inductors fabricated in-house~\cite{zhang_gate_2026} are wire-bonded to the three contacts of the device, with resonance frequencies indicated in the schematic. The superconductor is grounded through a bias tee. DC voltages can be applied to the normal contacts. In this configuration, reflectometry enables readout of the qubit via the LC resonator connected to the superconductor. Resonators connected to the normal leads are not used in this work.

\subsection{Device tune-up}\label{Methods_tuneup}
For the initial tune-up, we first characterize the device in the transport regime, with tunnel coupling to the normal leads. Following the procedure described in Ref.~\cite{zatelli_robust_2024}, we tune both Kitaev chains to a Majorana sweet spot. The device is then pinched off from the leads, after which all remaining tuning is performed using the signal reflected from the resonator connected to the superconductor.

The first step in the pinched-off regime is to characterize the coupling between each Kitaev chain and the coupler QD. This coupling is estimated by detuning one chain away from resonance and focusing on the subsystem formed by the coupler QD and the other chain. With this chain tuned on resonance, the coupler QD is pulsed while varying both the pulse amplitude and ramp time as shown in Fig.~\ref{fig:ed_rampduration}f,j. This measurement provides an estimate of both the resonance condition of the coupler QD and of the tunnel coupling strength, which is inferred from the ramp time at which the transition from diabatic to adiabatic ramps occurs. The tunnel barrier between the coupler QD and the on-resonance chain is then adjusted such that this transition occurs for ramp times below $\SI{10}{\nano\second}$. Pinching off the tunnel barrier reduces the coupling and shifts the transition to longer ramp times, whereas opening the barrier increases the coupling and shifts the transition to shorter ramp times. This procedure is performed independently for both chains, with the aim of obtaining approximately symmetric couplings to the coupler QD.

Once these couplings have been tuned, further optimization is performed using pulsed charge-stability diagrams (Fig.~\ref{fig:1}f,g, Methods~\ref{Methods_pulsedCSD}) and conditional-probability maps as a function of the left and right hybrid-gate voltages (Fig.~\ref{fig:3}c). The pulsed charge-stability diagrams are measured in both the global even and global odd parity manifold and are used to calibrate the pulse amplitudes required to bring the QDs into resonance. The hybrid-gate scans are used to identify the poor man's Majorana sweet spot conditions. As a final check, after measuring the pulsed charge-stability diagrams and the hybrid-gate scans, Rabi oscillations are measured in the global even and global odd parity manifolds at the identified sweet spot. A successful tuning is indicated by oscillations with similar frequencies and contrast in both manifolds, together with a decay of the conditional probabilities toward $0.5$.

This tuning procedure is used for all measurements shown in the main text. Since most two-dimensional measurements take between one and four hours, the device is retuned before each two-dimensional sweep to compensate for drift in the quantum-dot chemical potentials and hybrid-gate sweet spot conditions. The measurements are acquired in the following order: coupler-dot detuning versus pulse duration, shown in Fig.~\ref{fig:2}a and Fig.~\ref{fig:3}a; single-dot detuning maps, shown in Fig.~\ref{fig:4}a-d; double-dot detuning maps, shown in Fig.~\ref{fig:4}e-h; hybrid-gate detuning maps, shown in Fig.~\ref{fig:3}c-j; and finally the out-of-plane magnetic field sweeps, shown in Fig.~\ref{fig:ed_detunings}o and Fig.~\ref{fig:2}e. For the magnetic field sweep shown in Fig.~\ref{fig:ed_detunings}o, the device is retuned to the Majorana sweet spot at $\SI{10.5}{\milli\tesla}$ and $\SI{15.6}{\milli\tesla}$ to compensate for shifts in the chemical potentials of the QDs and Andreev bound states due to Zeeman field. For the magnetic field sweep shown in Fig.~\ref{fig:2}e, the device is retuned at $\SI{9.5}{\milli\tesla}$, $\SI{10.5}{\milli\tesla}$, $\SI{11.5}{\milli\tesla}$, $\SI{12.5}{\milli\tesla}$, $\SI{13.5}{\milli\tesla}$, $\SI{14.5}{\milli\tesla}$, $\SI{15.5}{\milli\tesla}$, $\SI{16.5}{\milli\tesla}$, and $\SI{17.5}{\milli\tesla}$.

Between the measurements shown in Fig.~\ref{fig:ed_detunings}o and Fig.~\ref{fig:2}e, a phase jump was observed. Before this jump, an out-of-plane magnetic field of $\SI{15.6}{\milli\tesla}$ corresponded to $\varphi \approx 0$. After the jump, the magnetic-field dependence of the oscillation frequency was shifted by approximately $\SI{2.5}{\milli\tesla}$. In addition, before measuring the data shown in Fig.~\ref{fig:2}e, a charge jump occurred, after which the device displayed increased instability.

The data shown in Fig.~\ref{fig:ed_pulsedCSDsQDC} and Fig.~\ref{fig:ed_reproduction} were acquired in a different cooldown before the measurement protocol for the global odd parity manifold had been developed. For this dataset, the tuning therefore relied on pulsed charge-stability diagram measurements in the global even parity manifold, together with fine-tuning of the hybrid-gate voltages. This procedure confirms the degeneracy condition within the global even parity manifold. The degeneracy between the two parity manifolds, however, could not yet be assessed for this dataset.

\subsection{Readout calibration}\label{Methods_readout}
Single-shot readout of the qubit is performed by measuring the reflected signal $\Delta\SM$ of the RF resonator connected to the superconducting lead, which senses the parity-dependent quantum capacitance of the minimal Kitaev chains~\cite{van_loo_single-shot_2026}. 

During readout, the coupler QD is detuned far off-resonance. Because the inter-chain coupling decreases only algebraically with the coupler gate voltage $\VC$, a residual coupling remains that can induce incoherent hopping between the chains and shorten their individual parity lifetime. When this lifetime becomes comparable to the single-shot readout time, parity switching during the measurement averages the quantum-capacitance signals and prevents reliable single-shot readout.

To suppress this effect, the two chains are detuned during readout, introducing an energy splitting between their states corresponding to an effective $Z$ term that exceeds the residual inter-chain coupling. The detuning is applied antisymmetrically within each chain to preserve sensitivity to the quantum capacitance. However, it also lifts the degeneracy between even and odd parity states, making the odd state energetically favorable and thereby reducing the lifetime of the even-parity state.
The detuning is therefore optimized to balance these competing effects: it must be large enough to suppress residual inter-chain coupling, while remaining small enough to maintain a sufficiently long lifetime of the even-parity state to achieve adequate readout contrast.
Single-shot readout is performed by integrating the reflectometry signal for $\SI{500}{\micro s}$, much faster than the shortest parity lifetime $\gtrsim \SI{8}{\milli s}$ (Fig.~\ref{fig:ed_readout}).

\subsection{Pulse details}\label{Methods_pulses}
A schematic of a typical pulse sequence is shown in Fig.~\ref{fig:1}e and Fig.~\ref{fig:ed_rampduration}a. Each sequence follows an identical protocol: it begins with a single-shot readout, followed by voltage pulses applied to the quantum dots, and concludes with a final single-shot readout. A waiting time of $\SI{50}{\micro\second}$ is inserted between the voltage pulses and the readout events. This waiting time is chosen such that, for measurements in the global odd parity manifold, the pulse applied to the parity-switching QD does not overlap with the readout events. By pulsing the parity-switching QD across charge degeneracy between the two readout events, the system is mapped into the global odd parity manifold during the pulsed operations, while readout remains in the global even parity manifold. The same waiting time is used for measurements in the global even parity manifold, except for those in Fig.~\ref{fig:ed_pulsedCSDsQDC}, Fig.~\ref{fig:ed_rampduration}f,j, and Fig.~\ref{fig:ed_reproduction}, where we used a waiting time of $\SI{5}{\micro s}$.

The voltage pulses applied to the quantum dots are generated by the Presto unit and applied simultaneously through RF lines. For the pulsed experiments presented in the main text and Extended Data, a ramp time of $\SI{25}{\nano\second}$ is used. The relative hardware delay between the different RF pulse lines is estimated to be $\lesssim \SI{2}{\nano\second}$ and is neglected, as it is short compared to the applied ramp times. To sweep the pulse duration with a resolution finer than the native $\SI{2}{\nano\second}$ time grid of the Presto unit, we follow Ref.~\cite{wang_operating_2024}. The pulse amplitudes shown in the main text and Extended Data correspond to the calibrated amplitudes arriving at the gates, rather than to the voltages output by the Presto.

The pulse applied to the parity-switching QD is generated by an arbitrary waveform generator built in-house and applied through a DC line, as this plunger gate was not wire-bonded to an RF line. This pulse has a ramp time of $\SI{200}{\nano\second}$ and a flat-top duration of $\SI{60}{\micro\second}$, with the delivered waveform bandwidth-limited by the RC filters on the DC line (see Fig.~\ref{fig:ed_measurementsetup}).

The pulse amplitudes and DC gate voltages shown in the main text and Extended Data are reported as virtual gate voltages that include cross-capacitance corrections. When pulsing the QDs, we compensate for cross-capacitance between $\VLR$, $\VC$, and $\VRL$. Cross-capacitance between $\VLL$ and $\VLR$, and between $\VRL$ and $\VRR$, is negligible due to screening by the superconductor. For measurements in the global odd parity manifold, we compensate for cross-capacitance on $\VRR$ from the pulse applied to $\VPS$. Additionally, we compensate for cross-capacitance from the hybrid gates to the nearest QD plunger-gates, $\VLL$, $\VLR$ for $\VLH$, and $\VRL$, $\VRR$ for $\VRH$. For the data shown in Fig.~\ref{fig:ed_reproduction}, we also apply corrections to $\VRL$ and $\VLR$ to compensate for cross-capacitance from $\VLH$ and $\VRH$, respectively.
In addition to cross-capacitance corrections, we characterized the interdot charging energy between \QDPS and \QDRR, which shifts the chemical potential of \QDRR during measurements in the global odd parity manifold. This shift is accounted for by calibrating the pulse amplitudes required to bring the quantum dots into resonance separately in the two global parity manifolds using pulsed charge-stability diagrams.

Two different implementations of this measurement protocol were used. In the first implementation, each individual repetition of the protocol consisted of an initial readout, a voltage-pulse sequence, and a final readout. This implementation was used only for the data shown in Fig.~\ref{fig:ed_pulsedCSDsQDC}, Fig.~\ref{fig:ed_rampduration}, and Fig.~\ref{fig:ed_reproduction}. In the later implementation, the final readout of one repetition was reused as the initial readout for the next repetition to reduce measurement time overhead. For most Rabi measurements shown in the main text and Extended Data, each data point was averaged over 2000-4000 repetitions.

Between measurements, an off-resonant drive at $\SI{580}{\mega\hertz}$ is applied to the RF input line to improve the stability of the reflected signal. Without this drive, a transient drift in the reflected signal is observed immediately after the readout tone is switched on. This drift stabilizes on a timescale of hundreds of milliseconds, depending on the applied readout drive amplitude. We attribute this behavior to heating of the attenuators of the readout line or of the directional coupler, with the associated timescale corresponding to the time required to reach a steady state. The off-resonant drive suppresses this transient response by keeping the readout line close to a steady state between measurements. It is switched off before the acquisition of each data point, remains off throughout all repetitions over which the data point is averaged, and is switched on again after the acquisition is completed.
For the data shown in Fig.~\ref{fig:ed_pulsedCSDsQDC} and Fig.~\ref{fig:ed_reproduction}, the off-resonant drive was not used. It was implemented during the acquisition of the remaining data to allow for a larger readout drive amplitude. 

\subsection{Rabi oscillation fit details}\label{Methods_fit}
Rabi oscillations were fitted to an exponentially damped cosine,
\begin{equation}
P(t)=A\exp\left(-\frac{t}{T_2^{\mathrm{Rabi}}}\right)
\cos\left(2\pi f t+\phi\right)+C,
\end{equation}
where $A$ is the oscillation amplitude, $T_2^{\mathrm{Rabi}}$ the decay time, $f$ the oscillation frequency, $\phi$ the phase, and $C$ a constant offset. In Fig.~\ref{fig:2} and Fig.~\ref{fig:3}, only data with $V_{\mathrm{C}}<\SI{11.25}{\milli V}$ were included in this fit, as the oscillations become too rapid at higher pulse amplitudes to be reliably sampled with the experimental time resolution. Initial frequency estimates were obtained from the Fourier spectrum of each trace, after which nonlinear least-squares optimization was performed using multiple initial guesses for the phase, oscillation frequency, and decay time to improve convergence and avoid local minima.
To this end, we used the function \texttt{curve\_fit} from the \texttt{scipy.optimize} Python library, setting its \texttt{absolute\_sigma} argument to \texttt{False}. Uncertainties on single data points have been assumed to be uniform across each dataset.
To improve numerical conditioning, the time axis was internally normalized during optimization and the fitted parameters were subsequently rescaled to the original physical units. The extracted oscillation frequencies were fitted to the inverse-detuning model
\begin{equation}
f=\frac{K}{V_\mathrm{0}-V_{\mathrm{C}}},
\end{equation}
where $V_{\mathrm{C}}$ is the coupler pulse amplitude and $V_\mathrm{0}$ is the fitted resonance position.

\subsection{Numerical simulations}\label{methods:numerical_simulations}

We numerically model the system presented in Fig.~\ref{fig:1} with the Hamiltonian

\begin{equation} \label{model:pmm_hamiltonian}
\begin{split}
    H &= H_\mathrm{L} + H_\mathrm{R} + H_\mathrm{C} + H_\mathrm{T}, \\
    H_\alpha &= \sum_i \mu_{\alpha i}c_{\alpha i}^\dagger c_{\alpha i} + t_\alpha c_{\alpha 1}^\dagger c_{\alpha 2} + \Delta_\alpha e^{i\varphi_\alpha}c_{\alpha 1}^\dagger c_{\alpha 2}^\dagger + \text{h.c.},\\
    H_\mathrm{C} &= \muC c_\mathrm{C}^\dagger c_\mathrm{C}, \quad
    H_\mathrm{T} = \tLC c_\mathrm{L2}^\dagger c_\mathrm{C} + \tCR c_\mathrm{C}^\dagger c_\mathrm{R1} + \text{h.c}.
    \end{split}
\end{equation}  

Here $H_{\alpha}$ for $\alpha=\mathrm{L}, \mathrm{R}$ is a two-site Kitaev chain Hamiltonian, with $\mu_{\alpha i}$ the on-site potential on site $i$ in chain $\alpha$.  $t_\alpha$ and $\Delta_\alpha$ are the ECT and CAR amplitudes mediated by Andreev bound states in the hybrid segment~\cite{liu_tunable_2022, bordin_tunable_2023}. 
We include the superconducting phase $\varphi_\alpha$ via $\varphi_\mathrm{L}=0$ and $\varphi_\mathrm{R}=\varphi$.
$\muC$ is the on-site potential of the coupler QD, 
and $\tLC$ and $\tCR$ are the tunneling amplitudes to the coupler QD.

In the presence of proximity effect on the dots in the chain closest to the coupler QD, an effective superconducting pairing term is also possible between the dots in the chain and the coupler QD~\cite{bordin_probing_2026}. 
We have found that such a term does not qualitatively change the results, and thus focus only on a normal hopping term to the coupler QD.

To simulate the time evolution of the Majorana parity qubit, we consider the full Hamiltonian (Eq.~\ref{model:pmm_hamiltonian}) and solve the qubit dynamics numerically.
We consider the initial state $|\psi_0\rangle=|ee\rangle$ where two chains are initially decoupled ($\muC\rightarrow\infty$) and the coupler QD is unoccupied.
We then set a lower value of $\muC$ to induce coherent oscillations between the states $|ee\rangle$ and $|oo\rangle$.
During this pulse, the system evolves under the coupled Hamiltonian (Eq.~\ref{model:pmm_hamiltonian}), with time-evolution operator $U(t)=\exp(-iHt/\hbar)$.
The evolved state is $|\psi(t)\rangle=U(t)|\psi_0\rangle$.
We determine the return probability to the initial state, $P(t)=|\langle \psi_0|\psi(t)\rangle|^2$, as a function of the pulse duration $t$.

To include the effect of noise, we make use of the quasistatic approximation~\cite{ithier_decoherence_2005, paladino_review_2014}: the Hamiltonian parameters are assumed to vary slowly compared to the measurement time, and are held constant during the time evolution. We then numerically average the return probability over an ensemble, where every Hamiltonian parameter is sampled independently from a Gaussian distribution around a mean value, with a given standard deviation (such as $\sigma_{\mu_{\alpha i}}$ around mean $\mu_{\alpha i}$).

The numerical simulation results are shown in Fig.~\ref{fig:ed_theory_even} and Fig.~\ref{fig:ed_theory_odd}, while the simulation parameters are reported in Table~\ref{tab:sim}. The chemical-potential intervals used in the simulations are chosen to approximately match the experimental data, assuming realistic plunger-gate lever arms ($\alpha_\mathrm{C} \approx 0.1$ for the coupler QD, $\alpha_i \approx 0.03$ for the QDs belonging to the chains). The standard deviations of the Gaussian noise on each simulation parameter are chosen to qualitatively reproduce the experimental data in Fig.~\ref{fig:ed_detunings} and Fig.~\ref{fig:ed_detunings_odd}. No fitting procedure is performed.

\begin{table}[h!]
\centering
\caption{Parameters and quasistatic Gaussian noise used for the full-model detuning simulations. Energies and standard deviations ($\sigma$) are in $\mu\mathrm{eV}$, except $\varphi$ in radians.}
\label{tab:detuning_simulation_parameters}
\begin{tabular}{cccccccccc}
\hline
 & $\tL$ & $\tR$ & $\DeltaL$ & $\DeltaR$ & $\varphi$ & $\muC$ & $\mu_{\alpha i}$ &$\tLC$ & $\tCR$ \\
\hline
Value 
& $20$ & $20$ & $20$ & $20$ & $0$ & $130$ & $0.0$ & $7.5$ & $7.5$ \\
$\sigma$
& $0.07$ & $0.07$ & $0.07$ & $0.07$ & $0$ & $4.5$ & $1.5$ & $0.1$ & $0.1$ \\
\hline
\end{tabular}
\label{tab:sim}
\end{table}

\subsection{Effective low-energy Hamiltonian for imperfect Majorana states}\label{methods:effective_full_hamiltonian}

Eqs.~\eqref{eq:hamiltonian_even} and \eqref{eq:hamiltonian_odd} of the main text state the effective qubit Hamiltonian when only nearest-neighbor couplings are included. To go beyond this approximation, we now consider an effective low-energy Hamiltonian including all possible coupling terms:
\begin{equation}\label{eq:eff_majorana_ham}
\begin{split}
    H_\text{eff} &= i\varepsilon_{12}  \gamma_1 \gamma_2 +  i \varepsilon_{34} \gamma_3 \gamma_4 + i \varepsilon_{23} \gamma_2\gamma_3 \\
    & + i\varepsilon_{24} \gamma_2\gamma_4+ i\varepsilon_{13} \gamma_1\gamma_3+ i\varepsilon_{14} \gamma_1\gamma_4. 
\end{split}
\end{equation}
Here, $\gamma_i$ are the four Majorana bound states forming the low-energy subspace of the left and right chains (see Fig.~\ref{fig:1} of the main text). The first three terms involving $\varepsilon_{12}$, $\varepsilon_{34}$, and $\varepsilon_{23}$ are the Majorana couplings considered in the main text. In contrast, $\varepsilon_{13}$, $\varepsilon_{24}$, and $\varepsilon_{14}$ are ``unwanted'' Majorana couplings, associated with longer-range overlaps. In this and the following sections, unlike in the main text, the superconducting phase dependence is absorbed into the couplings $\varepsilon_{ij}$.

The Hamiltonian of Eq.~\eqref{eq:eff_majorana_ham} can be written as an effective qubit Hamiltonian by defining $\left|0\right>$ and $\left|1\right>$ states in both the even and odd manifolds:
\begin{equation}
\begin{split}
    \text{global even parity:}& \left|0\right>_e = \left|ee\right>, \left|1\right>_e = \left|oo\right>,\\
    \text{global odd parity:}& \left|0\right>_o = \left|eo\right>, \left|1\right>_o = \left|oe\right>.
\end{split}
\end{equation}
Defining complex fermions as
\begin{equation}
f_L = \frac{\gamma_1 - i \gamma_2}{2},\quad f_R = \frac{\gamma_3 - i \gamma_4}{2},
\end{equation}
with $f_\alpha \left|e_\alpha\right> = 0$ and $f_\alpha \left|o_\alpha\right> = \left|e_\alpha\right>$ for $\alpha=\mathrm{L,R}$, we thus find in the even manifold
\begin{equation}\label{eq:qubit_ham_even}
    H_\text{even} = (\varepsilon_{12} + \varepsilon_{34}) Z + (\varepsilon_{23} +\varepsilon_{14}) X +(\varepsilon_{13} - \varepsilon_{24})Y.
\end{equation}
In the odd manifold, we obtain
\begin{equation}\label{eq:qubit_ham_odd}
    H_\text{odd} = (\varepsilon_{12} - \varepsilon_{34}) Z + (\varepsilon_{23} -\varepsilon_{14}) X +(\varepsilon_{13} + \varepsilon_{24})Y.\end{equation}
Compared to the idealized model of the main text, unwanted Majorana couplings modify the effective $X$ coupling and generate an additional $Y$ coupling. Since these corrections enter differently in Eqs.~\eqref{eq:qubit_ham_even} and \eqref{eq:qubit_ham_odd}, the transverse qubit coupling differs between even and odd manifolds when unwanted Majorana couplings are present. Conversely, similar Rabi frequencies in the even and odd manifolds indicate that unwanted Majorana couplings are small.

When the effective Hamiltonian in Eq.~\eqref{eq:eff_majorana_ham} has small unwanted Majorana couplings, it may be tempting to associate this finding with well-localized Majorana states.
However, this effective Hamiltonian can represent any system with two weakly coupled subsystems, each with nearly degenerate even and odd ground states~\cite{samuelson_quantifying_2026}. 
For example, quasi-Majorana states could yield a similar Hamiltonian with little unwanted Majorana coupling, despite the effective Majorana components not spatially separated in this case~\cite{vuik_reproducing_2019}. 
Hence, any conclusions obtained from comparing to this effective Hamiltonian can only make statements about effectively decoupled Majorana states, but not whether these are spatially separated, simply because the effective Hamiltonian \eqref{eq:eff_majorana_ham} is agnostic of the underlying microscopic model at this level.
Therefore, a careful study of detuning individual chain parameters described by Eq.~\eqref{model:pmm_hamiltonian}, as is done in the main text, is necessary to get evidence for spatial separation.

The microscopic Kitaev Hamiltonian of Eq.~\eqref{model:pmm_hamiltonian} can be reduced to this effective Hamiltonian.
Specifically, we have~\cite{tsintzis_majorana_2024}:
\begin{equation}
\begin{split}
    \varepsilon_{12} &= \frac{\delta \epsL}{2}, \quad \varepsilon_{34} = \frac{\delta \epsR}{2},\\
    \varepsilon_{23} &= \frac{\tLR}{2} \cos\frac{\varphi}{2}, \quad  \varepsilon_{14} = \frac{\tLR}{2}  \lambdaL \lambdaR \cos\frac{\varphi}{2}, \\
    \varepsilon_{13} & = - \frac{\tLR}{2} \lambdaL \sin\frac{\varphi}{2},\quad \varepsilon_{24} = \frac{\tLR}{2} \lambdaR \sin\frac{\varphi}{2}\,.
\end{split}
\end{equation}
Here $\delta \epsilon_\alpha= E_{\ket{e^{}_\alpha}} - E_{\ket{o^{}_\alpha}}$ ($\alpha=\mathrm{L,R}$) is the difference between the even ($E_{\ket{e^{}_\alpha}}$) and the odd ($E_{\ket{o^{}_\alpha}}$) ground state energies,
\begin{equation}\label{eq:kitaev_gs_even}
E_{\ket{e^{}_\alpha}}
=
\frac{\mu_{\alpha 1}+\mu_{\alpha 2}}{2}
-
\sqrt{
\left(
\frac{\mu_{\alpha 1}+\mu_{\alpha 2}}{2}
\right)^2
+
\Delta_\alpha^2
},
\end{equation}
and
\begin{equation}\label{eq:kitaev_gs_odd}
E_{\ket{o^{}_\alpha}}
=
\frac{\mu_{\alpha 1}+\mu_{\alpha 2}}{2}
-
\sqrt{
\left(
\frac{\mu_{\alpha 1}-\mu_{\alpha 2}}{2}
\right)^2
+
t_\alpha^2
}.
\end{equation}
$\tLR$ is the effective hopping between the two chains, mediated by the central dot.
$\lambda_\mathrm{L/R}$ is the wavefunction amplitude of the Majorana state localized towards the outer dots and close to the sweet spot given by
\begin{equation}\label{eq:majorana_wave_function_overlap}
\lambdaL = \frac{\muLL}{2 \tL} \text{ and }    \lambdaR = \frac{\muRR}{2 \tR}.
\end{equation}
The unwanted Majorana couplings are thus enabled by wavefunction overlap of the outer Majorana bound states with the inner dots. In particular, this wavefunction overlap only depends on the energies of the outer dots of each chain. Moreover, the phase $\varphi$ can be used to change which unwanted couplings are relevant.

In terms of microscopic parameters, the qubit Hamiltonians thus read
\begin{equation}\label{eq:qubit_kitaev_ham_even}
    H_\text{even} = \frac{\delta \epsL + \delta \epsR}{2} Z + \frac{\tLR}{2} \cos \frac{\varphi}{2} \left(1 + \frac{\muLL\muRR}{4\tL \tR}\right) X - \frac{\tLR}{2} \sin \frac{\varphi}{2} \left(\frac{\muLL}{2\tL} + \frac{\muRR}{2\tR}\right) Y
\end{equation}
and
\begin{equation}\label{eq:qubit_kitaev_ham_odd}
    H_\text{odd} = \frac{\delta \epsL - \delta \epsR}{2} Z+ \frac{\tLR}{2} \cos \frac{\varphi}{2} \left(1 - \frac{\muLL\muRR}{4\tL \tR}\right) X - \frac{\tLR}{2} \sin \frac{\varphi}{2} \left(\frac{\muLL}{2\tL} - \frac{\muRR}{2\tR}\right) Y.
\end{equation}

\subsection{Interpretation of pulsed charge-stability diagrams}\label{Methods_pulsedCSD}
In the main text, we use pulsed charge-stability diagrams to tune the Kitaev chains to the Majorana sweet spot (Fig.~\ref{fig:1}f,g). Here, we clarify the origin and interpretation of the lines observed in these diagrams.

We first consider the inner lines in the global even manifold. These lines arise when the $\ket{ee}$ and $\ket{oo}$ states are on resonance, i.e. $E_{\ket{ee}}=E_{\ket{oo}}$. Equivalently, in terms of individual chain splitting, this condition is $\varepsilon_{12}= - \varepsilon_{34}$ (see Eq.~\eqref{eq:hamiltonian_even}). When these two states are resonant, coupling between the chains results in enhanced parity mixing, as observed experimentally.

In a pulsed charge-stability diagram, the chemical potentials of one chain are varied while the other chain is kept fixed. As a result, the observed parity-mixing lines depend, in general, also on the chain that is not being detuned. For example, if we measure a pulsed charge-stability diagram of the right chain and the left chain has a finite splitting $\overline{\varepsilon}_{12}$, the parity-mixing lines do not correspond to even--odd degeneracies of the right chain. Instead, they occur where $\varepsilon_{34}=-\overline{\varepsilon}_{12}$.
The same description applies to the global odd manifold, with the exception that the resonance condition is $\varepsilon_{12}= \varepsilon_{34}$  (see Eq.~\eqref{eq:hamiltonian_odd}).

We illustrate this behavior in Fig.~\ref{fig:ed_PCSDs}. When all QDs are pulsed to charge degeneracy, the energy splittings of the two chains are $\varepsilon_{12} = (\tL - \DeltaL)/2$ and $\varepsilon_{34} = (\tR - \DeltaR)/2$.
Crossings are observed in both parity manifolds and for both chains only at the sweet spot, $\tL=\DeltaL$ and $\tR=\DeltaR$ (Fig.~\ref{fig:ed_PCSDs}g).
However, detuning the hybrid gate of a single chain (Fig.~\ref{fig:ed_PCSDs}d,f,h,j) opens avoided crossings in the pulsed charge-stability diagrams measured for both chains. These avoided crossings should not be interpreted as evidence that both chains are individually detuned from their sweet spots. Rather, the parity-mixing lines indicate resonances of the full two-chain system. A finite splitting in one chain therefore shifts the resonance condition probed when pulsing the other chain, making the effect of detuning one chain visible in both pulsed diagrams.
To ensure that the observed features reliably identify even--odd degeneracies of an individual chain like in typical charge-stability diagrams, it is therefore crucial to ensure that the other chain is tuned to degeneracy.

Importantly, a crossing in a single global-parity manifold is not sufficient to identify the sweet spot. Away from the sweet spot, the resonance condition can still be satisfied in only one manifold if the splittings of the two chains compensate each other. At charge degeneracy, this occurs when $\tL-\DeltaL=\DeltaR - \tR$ in the even manifold, and when $\tL-\DeltaL=\tR - \DeltaR$ in the odd manifold, as shown in Fig.~\ref{fig:ed_PCSDs}c,e,i,k. Therefore, an apparent crossing in one manifold only shows that the corresponding two-chain resonance condition is satisfied, but it does not imply $\tL=\DeltaL$ and $\tR=\DeltaR$. Instead, we emphasize that the chains are tuned to the Majorana sweet spot only when pulsed charge-stability diagrams of both the left and right chains show crossings in both parity manifolds, as shown in Fig.~\ref{fig:ed_PCSDs}g and Fig.~\ref{fig:ed_PCSDs_excited}. We verify this condition before every measurement presented in the main text.

Finally, we comment on the outer lines shown in the pulsed charge-stability diagrams of Fig.~\ref{fig:1}f,g and Fig.~\ref{fig:ed_PCSDs_excited}. For simplicity, we restrict our discussion to the case where both chains are tuned to the sweet spot, $\tL = \DeltaL$ and $\tR = \DeltaR$, and the left-chain QDs are tuned to zero chemical potential. Additional lines in the pulsed charge-stability diagram of the right chain appear when its ground state is on resonance with an excited state of the left chain, allowing them to be mixed by inter-chain coupling. If the chains are initialized in the $\ket{oo}$ state, the resonance condition is satisfied when $E_{\ket{oo}}=E_{\ket{e^+e}}$, where $\ket{e^+}$ denotes the first even excited state of the left chain. With the assumptions above, this condition simplifies to $-\varepsilon_{34} = 2\abs{\DeltaL}$. In other words, the splitting between even and odd ground states of the right chain must be equal to the excitation gap of the left chain. This condition can be satisfied only if $\varepsilon_{34}<0$, i.e. when the even ground state is lower in energy than the odd ground state. This is consistent with the line appearing only when the right-chain QDs are detuned symmetrically away from the sweet spot (top-right and bottom-left quadrant of Fig.~\ref{fig:1}f).
If the chains are initialized in $\ket{ee}$, the resonance condition is modified to $E_{\ket{ee}}=E_{\ket{o^+o}}$, where $\ket{o^+}$ denotes the first odd excited state of the left chain. This condition simplifies to $\varepsilon_{34} = 2\abs{\DeltaL}$. The opposite sign explains why the excited-state lines appear in different quadrants for different initial states (Fig.~\ref{fig:ed_PCSDs_excited}). The observations in the odd manifold can be understood analogously.

We emphasize that the pulsed charge-stability diagrams show conditional probabilities between the ground state configurations, namely between $\ket{ee}$ and $\ket{oo}$ in the global even parity manifold and between $\ket{eo}$ and $\ket{oe}$ in the global odd parity manifold. They do not directly show conditional probabilities involving excited states. In the final readout after the pulse sequence, the excited states are not detected, which we attribute to their fast relaxation~\cite{van_loo_single-shot_2026}. As a result, mixing between, for example, the $\ket{oo}$ ground state and the $\ket{e^{+}e}$ excited state is detected as a mixing between the $\ket{oo}$ and $\ket{ee}$ ground states.

\subsection{Theory of pulsed charge-stability diagrams in the general case}\label{methods:theory_of_pulsed_csd}

The arguments of the previous section can be extended into a mathematical description of the lines corresponding to the ground states in the observed pulsed charge-stability diagrams.
In these pulsed charge-stability diagrams, one chain remains unchanged, whereas both chemical potentials are varied in the other chain. We consider the case that the chemical potentials of the left chain, $\muLL$ and $\muLR$ are varied, whereas the splitting $\delta \epsR$ is constant. The condition for finding zero longitudinal splitting in Eqs.~\eqref{eq:qubit_ham_even} and \eqref{eq:qubit_ham_odd} reduces to
\begin{equation}
\delta \epsL + \delta \epsR = 0 \quad \text{in the even and}\quad
\delta \epsL -\delta \epsR = 0, \quad \text{in the odd manifold.}
\end{equation}
Thus, when the parameters of the right chain remain fixed, this reduces to solving for
\begin{equation}
    \delta \epsL = \epsilon,
\end{equation}
with $\epsilon = -\delta \epsR$ in the even manifold, and $\epsilon = \delta \epsR$ in the odd.

Using the microscopic expressions \eqref{eq:kitaev_gs_even} and \eqref{eq:kitaev_gs_odd} for the even and odd ground state energies, we then arrive at the conditions
\begin{equation}\label{eq:analytical_csd}
\begin{split}
    \muLR &= \frac{ \muLL \left(\tL^{2} - \DeltaL^{2} \right)
    \pm \left|{\epsilon}\right| \sqrt{\left(\muLL^{2} - \epsilon^{2} + \left(\DeltaL - \tL\right)^{2}\right) \left( \muLL^{2} - \epsilon^{2} + \left(\DeltaL + \tL\right)^{2}\right)}}{\muLL^{2} - \epsilon^{2}},\\
&\epsilon + \frac{\sqrt{4 \tL^{2} + \left(\muLL - \muLR\right)^{2}}}{2} \geq 0, \text{ and}\\
&\operatorname{sign}{\left(\DeltaL^{2} - \epsilon^{2} + \muLL \muLR - \tL^{2} \right)} = \operatorname{sign}{\left(\epsilon \right)}.
\end{split}
\end{equation}
A detailed derivation is given in the data repository.

When the right chain is already tuned to the sweet spot, $\epsilon=0$, these conditions reduce to
\begin{equation}
    \muLL \muLR = \tL^2 - \DeltaL^2.
\end{equation}
We show the lines of zero splitting as black solid lines in the $\muLL$-$\muLR$-plane in Fig.~\ref{fig:tuning_theory} for the cases $\tL<\DeltaL$ (CAR-dominated), $\tL=\DeltaL$ (sweet spot), and $\tL>\DeltaL$ (ECT-dominated).
These are equivalent to the charge-stability diagrams that have been measured in transport~\cite{dvir_realization_2023, zatelli_robust_2024, ten_haaf_two-site_2024}.

However, \emph{a priori} it is not known how well the other chain is tuned to zero. We also show the lines of zero splitting for the case of a finite $\epsilon=\SI{10}{\micro\electronvolt}$ as red dashed lines in Fig.~\ref{fig:tuning_theory}. These strongly resemble the true CAR to ECT transition we observe for $\epsilon=0$, but in this case the crossing of two lines of zero splitting does \emph{not} signify a Kitaev sweet spot as seen from the parameters of the plot. However, in the other parity manifold, $\epsilon = \SI{-10}{\micro\electronvolt}$. Hence, the lines of zero splitting in the other manifold will look very different for the same parameters, as observed also experimentally in Fig.~\ref{fig:ed_PCSDs}.

Fortunately, it is still possible to tune to $\muLL=\muLR=0$ by measuring only in a single parity manifold: for any value of $\epsilon$, the zero-splitting lines can only cross at $\muLL=\muLR=0$ as shown in two examples in Fig.~\ref{fig:tuning_theory}b. Mathematically, two zero-splitting lines can only cross at a critical point of $\delta \epsL$, with
\begin{equation}
    \frac{\partial}{\partial \muLL} \delta\epsL = \frac{\partial}{\partial \muLR} \delta\epsL = 0,
\end{equation}
which can only be fulfilled for $\muLL=\muLR=0$. Moreover, the value of the critical point is zero for $\epsilon = \left|{\DeltaL}\right| - \left|{\tL}\right|$. Details of this calculation are given in the data repository.

Thus, it is possible to use pulsed charge-stability diagrams in a single parity manifold to tune the chemical potentials in both chains to zero, $\mu_i\approx 0$ for $i=1,\ldots,4$. Then, the residual energy splitting in each chain is given by $\delta \epsilon_\mathrm{L/R} = \left|\Delta_\mathrm{L/R}\right|-\left|t_\mathrm{L/R}\right|$. This difference can be changed linearly to first order with the hybrid plunger gate. Hence, the longitudinal splitting in the even ($\delta \epsL + \delta \epsR$) and in the odd manifold ($\delta \epsL - \delta \epsR$) will have opposite slopes in the plane of the two hybrid plunger gates, as observed in Fig.~\ref{fig:3}.

\subsection{Consequences of imperfect Majorana states}\label{methods:consequences_imperfect_majoranas}

To study the consequences of imperfect Majorana states it is instructive to consider specific phase differences, $\varphi=0$ and $\varphi=\pi$.

At $\varphi=0$, unwanted Majorana overlap leads to different Rabi oscillation frequencies in the even and odd manifolds~\cite{tsintzis_majorana_2024}. These frequencies are $f_\text{even/odd} = \frac{1}{h} (\varepsilon_{23} \pm \varepsilon_{14})$, where the $+$ ($-$)-sign holds for the even (odd) manifold. Hence, the frequency difference shown in Fig.~\ref{fig:3}b can be expressed as
\begin{equation}
\frac{f_\text{even} - f_\text{odd}}{f_\text{even}+f_\text{odd}} = \frac{\varepsilon_{14}}{\varepsilon_{23}} = \frac{\muLL\muRR}{4 \tL \tR}.
\end{equation}
Since this frequency difference only depends on the product of both Majorana overlaps, it would be suppressed even if only one chain is at the sweet spot.
Hence, the observation of very small frequency differences in Fig.~\ref{fig:3}b strongly hints at well-decoupled Majorana states in at least one of the chains.
Further quantifying unwanted Majorana overlap from the measurement in Fig.~\ref{fig:3} is not possible without additional assumptions.

In contrast to $\varphi=0$, at phase difference $\pi$ the unwanted Majorana overlap enters linearly in the transverse coupling. 
In fact, a finite transverse qubit coupling is then \emph{only} possible if there is such a Majorana overlap.
Close to the sweet spot, the unwanted Majorana overlap depends to first order only on the detuning of the outermost dots of both chains, as seen from Eq.~\eqref{eq:majorana_wave_function_overlap}.

This is consistent with our experimental findings for the pulsed charge-stability diagrams at $\varphi\approx \pi$ in Figs.~\ref{fig:ed_pulsedCSDsQDC}i and j, where we see a clear difference between detuning the two dots in the right chain. When the outermost dot remains on resonance, and only the inner dot is detuned, we see a suppression of dephasing compared to the case $\varphi\approx 0$. This is consistent with a suppressed transverse coupling in this case. In contrast, when the outermost dot is detuned, we quickly see dephasing comparable to the case $\varphi \approx 0$. This indicates that the outermost dot detuning has enabled a finite transverse qubit coupling. Note that in Figs.~\ref{fig:ed_pulsedCSDsQDC}k-n the central dot is close to resonance, and the low-energy Hamiltonian in Eq.~\eqref{eq:eff_majorana_ham} is not valid. Hence the present arguments do not hold in this case.

At phase $\pi$, unwanted Majorana couplings determine the transverse qubit coupling linearly. 
This means that close to the sweet spot, noise can have a significant effect.
At the sweet spot, splitting of Majorana states in a two-site chain is quadratic in $\mu_{\alpha i}$.
Hence, to lowest order in chemical potential, we have
\begin{equation}
H_\text{even/odd} = H_\pm = \frac{\tLR}{2} \lambda_\pm Y,
\end{equation}
where the $+$ ($-$) sign corresponds to the even (odd) manifold, and $\lambda_\pm = \frac{\muLL}{2 \tL} \pm \frac{\muRR}{2 \tR}$.
The return probability is then given by
\begin{equation}
P(t) = \frac{1}{2} + \frac{1}{2} \cos\left(\frac{\tLR}{\hbar} \lambda_\pm t \right).
\end{equation}
When we consider the case of random Gaussian noise on the chemical potential on the outermost two dots with standard deviation $\sigma_{\muLL}$ and $\sigma_{\muRR}$ and mean 0, $\lambda_\pm$ is also a Gaussian variable with mean 0 and standard deviation $\sigma_\lambda = \sqrt{\sigma_{\muLL}^2/(2\tL)^2 + \sigma_{\muRR}^2/(2\tR)^2}$, independent of the sign.
In the quasistatic approximation, we thus find for the average return probability in both manifolds
\begin{equation}
\begin{split}
    \langle P(t) \rangle & = \frac{1}{2} + \frac{1}{2} \int_{-\infty}^{\infty} \mathrm{d}\lambda \frac{1}{\sqrt{2 \pi} \sigma_\lambda} e^{-\lambda^2/2\sigma_\lambda^2} \cos\left(\frac{\tLR}{\hbar} \lambda t \right)\\
    &= \frac{1}{2} +\frac{1}{2} e^{-\frac{1}{2}  \left(\frac{\sigma_{\muLL}^2}{(2 \tL)^2} + \frac{\sigma_{\muRR}^2}{(2 \tR)^2}\right) \left(\frac{\tLR t}{\hbar}\right)^2}\,.
\end{split} 
\end{equation}
Noise-induced unwanted Majorana coupling thus gives rise to dephasing at $\varphi=\pi$, even at the sweet spot of the two-site chain.
This could explain the dephasing seen in Fig.~\ref{fig:ed_reproduction}b.
In general, the absence of an oscillatory behavior at $\varphi=\pi$ is consistent with noise in the unwanted Majorana coupling larger than its mean value, and thus with well-decoupled Majorana states in both chains.

\subsection{Use of large language models}
Large language models, primarily ChatGPT 5.4 and 5.5 (OpenAI), were used to assist with copy editing and with coding tasks related to measurement-code development, data analysis, plotting, symbolic calculations, and numerical simulations. Outputs were reviewed and validated by the authors responsible for the corresponding aspects of the work, as specified in the Author Contributions statement. The authors take responsibility for the final content accordingly.

\clearpage
\bibliography{sn-bibliography}


\end{document}